\documentstyle[12pt,psfig,myplots,preprint]{aastex}
\begin{document}

\textwidth 6.5in
\textheight 8.5in
\topmargin 0.0in
\oddsidemargin 0.0in

\title{
Lunar\hskip 0.05in Outgassing,\hskip 0.05in Transient\hskip 0.05in Phenomena\hskip
0.05in \&\hskip 0.05in the\hskip 0.05in Return\hskip 0.05in to\hskip 0.05in
the\hskip 0.05in Moon
II: Predictions and Tests for Outgassing/Regolith Interactions}

\author{Arlin P.S.~Crotts and Cameron Hummels}
\affil{Department of Astronomy, Columbia University,
Columbia Astrophysics Laboratory,\\
550 West 120th Street, New York, NY 10027}

\begin{abstract}
We follow Paper I with predictions of how gas leaking through the lunar surface
could influence the regolith, as might be observed via optical Transient
Lunar Phenomena (TLPs) and related effects.
We touch on several processes, but concentrate on low and high flow rate
extremes, perhaps the most likely.
We model explosive outgassing for the smallest gas overpressure at
the regolith base that releases the regolith plug above it.
This disturbance's timescale and affected area are consistent with observed
TLPs; we also discuss other effects.

For slow flow, escape through the regolith is prolonged by low diffusivity.
Water, found recently in deep magma samples, is unique among candidate
volatiles, capable of freezing between the regolith base and surface,
especially near the lunar poles.
For major outgassing sites, we consider the possible accumulation of water ice.
Over geological time ice accumulation can evolve downward through the regolith.
Depending on gases additional to water, regolith diffusivity might be
suppressed chemically, blocking seepage and forcing the ice zone to expand to
larger areas, up to km$^2$ scales, again, particularly at high latitudes.

We propose an empirical path forward, wherein current and forthcoming
technologies provide controlled, sensitive probes of outgassing.
The optical transient/outgassing connection, addressed via Earth-based remote
sensing, suggests imaging and/or spectroscopy, but aspects of lunar outgassing
might be more covert, as indicated above.
TLPs betray some outgassing, but does outgassing necessarily produces TLPs?
We also suggest more intrusive techniques from radar to in-situ probes.
Understanding lunar volatiles seems promising in terms of resource exploitation
for human exploration of the Moon and beyond, and offers interesting scientific
goals in its own right.
Many of these approaches should be practiced in a
pristine lunar atmosphere, before significant confusing signals likely to
be produced upon humans returning to the Moon.

\end{abstract}

\medskip
\section{Introduction}

Transient lunar phenomena (TLPs or LTPs) are defined for the purposes of this
investigation as localized (smaller than a few hundred km across), transient
(up to a few hours duration, and probably longer than typical impact events -
less than 1s to a few seconds), and presumably confined to processes near the
lunar surface.
How such events are manifest is summarized by Cameron (1972).
In Paper I (Crotts 2008; see also Crotts 2009) we study the systematic behavior
(especially the spatial distribution) of TLP observations - particularly their
significant correlations
with tracers of lunar surface outgassing, and we are thereby motivated to
understand if this correlation is directly causal.

Numerous works have offered hypotheses for the physical cause of TLPs
(Mills 1970, Garlick et al.\ 1972a, b, Geake \& Mills 1977, Cameron 1977,
Middlehurst 1977, Hughes 1980, Robinson 1986, Zito 1989, Carbognani 2004,
Davis 2009), but we present a methodical examination of the influence of
outgassing, exploring quantitatively how outgassing might produce TLPs.
Furthermore, it seems likely that outgassing activity is concentrated in
several areas, which leads one to ask how outgassing might interact with and
alter the regolith presumably overlying the source of gas.
Reviews of similar processes exist but few integrate
Apollo-era data e.g., Stern (1999), Mukherjee (1975), Friesen (1975).

As the final version of this paper approached completion, several papers were
published regarding the confirmed discovery of hydration of the lunar regolith.
Fortunately, we deal here with the special effects of water on lunar regolith
and find that many of our predictions are borne out in the recently announced
data.
We will deal with this explicitly in \S5.

Several experiments from Apollo indicate that gas is produced in
the vicinity of the Moon, even though these experiments disagree on the total
rate:
1) LACE (Lunar Atmosphere Composition Experiment on {\it Apollo 17}), $\sim$0.1
g s$^{-1}$ over the entire lunar surface (Hodges et al.~1973, 1974);
2) SIDE (Suprathermal Ion Detector Experiment on {\it Apollo 12, 14, 15}),
$\sim$7 g s$^{-1}$ (Vondrak et al.~1974);
3) CCGE (Cold Cathode Gauge Experiment on {\it Apollo 12, 14, 15}), $\la $60 g
s$^{-1}$ (Hodges et al.~1972).
These measurements not only vary by more than two orders of magnitude but also
in assayed species and detection methods.
LACE results here applies only to neutral $^{40}$Ar, $^{36}$Ar and $^{20}$Ne.
By mass $^{40}$Ar predominates.
SIDE results all relate to ions, and perhaps include a large contribution from
molecular species (Vondrak et al.\ 1974).
CCGE measures only neutral species, not easily distinguishing between them.

The LACE data indicate $^{40}$Ar episodic outgassing on timescales of a few
months or less (Hodges \& Hoffman 1975), but resolving this into faster
timescales is more ambiguous.
In this discussion we adopt the intermediate rate (SIDE), about 200 tonne
y$^{-1}$ for the total production of gas, of all species, ionized or neutral.
The LACE is the only instrument to provide compositional ratios, which also
include additional, rarer components in detail.
We will use these ratios and in some cases normalize them against the SIDE
total.

Much of the following discussion is only marginally sensitive to the actual
composition of the gas.
For many components of molecular gas at the lunar surface, however, there is a
significant possible contribution from cometary or meteoritic impacts, and a
lesser amount from solar wind/regolith interactions.
The influx of molecular gas from comets and meteorites are variously estimated,
usually in the range of tonnes or tens of tonnes per year over the lunar
surface (see Anders et al.\ 1973, Morgan \& Shemansky 1991).
Cometary contributions may be sporadically greater (Thomas 1974).
Except for H$_2$, solar wind interactions (Mukhergee 1975) provide only a small
fraction of the molecular concentration seen at the surface (which are only
marginally detected: Hoffman \& Hodges 1975).
There is still uncertainty as to what fraction of this gas is endogenous.
Current data do not succeed in resolving these questions, but we will return to
consider them later in the context of gas seepage/regolith interactions.

In this paper we consider various effects of outgassing through the regolith,
and find the most interesting simple effect occurs when the flow is high enough
to cause disruption of the regolith by an explosion to relieve pressure (\S 2),
which we compare to fluidization.
Another interesting effect occurs when the gas undergoes a phase change while
passing through the regolith (\S 3), which seems to apply only to water vapor.
This leads primarily to the prediction of the likely production of subsurface
ice, particularly in the vicinity of the lunar poles.
These effects suggest a variety of observational/experimental approaches, which
we summarize in \S 4.
In \S 5 we discuss the general implications of these findings, with specific
suggestions as to how these might guide further exploration, particularly in
respect to contamination by anthropogenic volatiles.
We also discuss the relevance of the predictions in \S 4 to very recent
discoveries regarding lunar regolith hydration.

First, let us make a few basic points about outgassing and the regolith.
One can easily picture several modes in which outgassing volatiles might 
interact with regolith on the way to the surface.
These modes will come into play with increasing gas flow rate and/or
decreasing regolith depth, and we simply list them with mneumonic labels along
with descriptions:

\noindent
1) choke: complete blockage below the regolith, meaning that any chemistry or
phase changes occur within the bedrock/megaregolith;

\noindent
2) seep: gas is introduced slowly into the regolith, essentially molecule by
molecule;

\noindent
3) bubble: gas is introduced in macroscopic packets which stir or otherwise
rearrange the regolith (such as ``fluidization'' e.g., Mills 1969);

\noindent
4) gulp: gas is introduced in packets whose adiabatic expansion deposits
kinetic energy into regolith and cools the gas, which therefore might even
undergo a phase change;

\noindent
5) explode: gas is deposited in packets at base of the regolith leading to an
explosion; and

\noindent
6) jet: gas simply flows into the vacuum at nearly sound speed with little
entrained material.

\noindent
While the intermediate processes might prove interesting, the extreme cases are
probably more likely to be in effect and will receive more of our attention.
In fact choking behavior might lead to explosions or geysers, when the pressure
blockage is released.
Since these latter two processes involve primarily simple hydrodynamics (and
eventually, newtonian ballistics), we will consider them first, and how they
might relate to TLPs.

\section{Explosive Outgassing}

If outgassing occurs at a rate faster than simple percolation can sustain, and
where regolith obstructs its path to the surface, the accumulation of the gas
will disrupt and cause bulk motion of the intervening regolith.
The outgassing can lift the regolith into a cloud in the temporary atmosphere
caused by the event.
The presence of such a cloud has the potential to increase the local albedo
from the perspective of an outside observer due to increased reflectivity and
possible Mie scattering of underlying regolith.
Additionally, volatiles buried in the regolith layer could become entrained in
this gas further changing the reflective properties of such a cloud.
Garlick et al.\ (1972b) describe fluidization of lunar regolith, in which dust
is displaced only temporarily and/or over small distances compared to ballistic
trajectories, but we will assume that we are dealing with more rapid changes.

\subsection{Model Explosion}

Let us construct a simple model of explosive outgassing through the lunar
surface.
For such an event to occur, we assume a pocket of pressurized gas builds at the
base of the regolith, where it is delivered by transport through the
crust/megaregolith presumably via channels or cracks, or at least faster
diffusion from below.
Given a sufficient flow rate (which we consider below), gas will accumulate at
this depth until its internal pressure is sufficient to displace the overlying
regolith mass, or some event releases downward pressure e.g., impact,
moonquake, incipient fluidization, puncturing a seal, etc.
We can estimate the minimal amount of gas alone required to cause explosive
outgassing by assuming that the internal energy of the buried gas is equal to
the total energy necessary to raise the overlying cone of regolith to the
surface.
This ``minimal TLP'' is the smallest outgassing event likely to produce
potentially observable disruption at a new site, although re-eruption through
thinned regolith will require less gas.

We consider the outgassing event occurring in two parts illustrated in Figure 1.
Initially, the gas bubble explodes upward propelling regolith with it until it
reaches the level of the surface; we assume that the plug consists of a cone
of regolith
within 45$^\circ$ of the axis passing from the gas reservoir to the surface,
normal to the surface.
Through this process the gas and regolith become mixed, and we assume they now
populate a uniform hemispherical distribution of radius $r \approx 14$~m on the
surface.
At this point, the gas expands into the vacuum and drags the
entrained regolith outward until the dust cloud reaches a sufficiently small density to
allow the gas to escape freely into the vacuum and the regolith to fall
eventually to the surface.
We consider this to be the ``minimal TLP'' for explosive outgassing, as there
is no additional reservoir that is liberated by the event beyond the minimum
to puncture the regolith.
One could also imagine triggering the event by other means, many of which might
release larger amounts of gas other than that poised at hydrodynamical
instability.

For the initial conditions of the first phase of our model, we assume the gas
builds up at the base of the regolith layer at a depth of 15~m (for more
discussion of this depth, see \S 3).
We set the bulk density of the regolith at $1.9$ g cm$^{-3}$ (McKay et
al.\ 1991), thereby setting the pressure at this depth at 0.45 atm.
Because of the violent nature of an explosive outgassing, we assume that the
cone of dust displaced will be 45$^\circ$ from vertical (comparable to the
angle of repose for a disturbed slope of this depth: Carrier, Olhoeft \&
Mendell 1991).
The mass of overlying regolith defined by this cone is $m= 7 \times 10^6$ kg.

In order to determine the mass of gas required to displace this regolith cone,
we equate the internal energy of this gas bubble with the potential energy ($U
= mgh$, where $g = 1.62$ m s$^{-2}$) required to lift the cone of regolith a
height $h = 15$~m to the surface, requiring 47,000 moles of gas.
Much of the gas found in outgassing events consists of $^4$He, $^{36}$Ar and
$^{40}$Ar (see \S 1), so we assume a mean molar mass for the model gas of $\mu
\approx 20$ g mol$^{-1}$, hence 940~kg of gas is necessary to create an
explosive outgassing event.
The temperature at this depth is $\sim0^\circ$C (see \S 3), consequently
implying an overall volume of gas of 2400 m$^3$ or a sphere 8.3 m in radius.
What flow rate is needed to support this?

Using Fick's diffusion law, $j ~ = ~ -K ~ dn/dz$, where the gas number density
$n = 1.1 \times 10^{19}$ cm$^{-3}$ is taken from above, and drops to zero
through 15~m of regolith in $z$.
The diffusivity $K$ is 7.7 and 2.3 cm$^2$ s$^{-1}$ for He and Ar, respectively,
in the Knudsen flow regime for basaltic lunar soil simulant (Martin et
al.\ 1973\footnote{For the relatively nonreactive gases He and Ar, the
diffusivity is proportional to $(T/\mu)^{1/2}$, where $T$ is the absolute
temperature.
The sticking time of the gas molecules, or heat of absorption, becomes
significant if the gas is more reactive or the temperature is reduced.
Unfortunately we find no such numbers for real regolith, although we discuss
realistic diffusivities for other gases below.}), so we adopt
$K = 5$ cm$^2$ s$^{-1}$ for our assumed He/Ar mixture.
(For any other gas mixture of this molecular weight, $K$ would likely be
smaller; below we also show that $K$ tends to be lower for real regolith.)~
Over the area of the gas reservoir, this implies a mass leakage rate of
2.8~g~s$^{-1}$, or $\sim$40\% of the total SIDE rate.
With the particular approximations made about the regolith diffusivity, this is
probably near the upper limit on the leakage rate.
At the surface of the regolith, this flow is spread to a particle flux of only
$\sim10^{16}$ cm$^{-2}$, which presumably causes no directly observable optical
effects.
The characteristic time to drain (or presumably fill) the reservoir is 4 d.

The second phase of the simulation models the evolution of dust shells and expanding
gas with a spherically symmetric 1D simulation centered on the explosion point.
The steps of the model include:
1) The regolith is divided into over 600 bins of different mean particle
size.  These bins are logarithmically spaced over the range $d = 8$~mm to $d =
0.02\mu$m according to the regolith particle size distribution from sample 72141,1
(from McKay et al.~1974).
The published distribution for sample 72141,1 only goes to 2$\mu$m, but other
sources (Basu \& Molinaroli 2001) indicate a component extending below 2$\mu$m, so
we extend our size distribution linearly from 2$\mu$m to 0$\mu$m.
Furthermore, we assume the regolith particles are spherical in shape and do not
change in shape or size during the explosion;
2) To represent the volume of regolith uniformly entrained in the gas, we create a
series of 1000 concentric hemispherical shells for each of the different particle
size bins (i.e., roughly 600,000 shells).
Each of these shells is now independent of each other and totally dependent on the
gas-pressure and gravity for motion;
3) We further assume that each regolith shell remains hemispherical throughout the
simulation.
Explicitly, we trace the dynamics of each shell with a point particle,
located initially 45 degrees up the side of the shell;
4) We calculate the outward pressure of the gas exerted on the dust shells.
The force from this pressure is distributed among different shells of regolith
particle size weighted by the total surface area of the grains in each shell.
We calculate each shell's outward acceleration and consequently integrate their
equations of motion using a timestep $dt = 0.001$ s;
5) We calculate the diffusivity of each radial shell (in terms of the ability
of the gas to move through it) by dividing the total surface area of all dust
grains in a shell by the surface area of the shell itself (assuming grains
surfaces to be spherical);
6) Starting with the largest radius shell, we sum the opacities of each shell
until we reach a gas diffusive opacity of unity.   Gas interior to this radius
cannot ``see'' out of the external regolith shells and therefore remains trapped.
Gas outside of this unit opacity shell is assumed to escape and is dropped from the
force expansion calculation.  Dust shells outside the unit opacity radius are now
assumed to be ballistic;
7) We monitor the trajectory of each dust shell (represented by its initially
45$^\circ$ particle) until it drops to an elevation angle of 30$^\circ$ (
when most of the gas is expanding above the particle), at which time
this particle shell is no longer supported by the gas, and is dropped from the
gas-opacity calculation;
8) An optical opacity calculation is made to determine the ability of an
observer to see the lunar surface when looking down on the cloud.
We calculate the downward optical opacity (such as from Earth) by dividing the
total surface area of the dust grains in a shell by the surface area of the
shell as seen from above ($\pi r^2$).
Starting with the outmost dust shell, we sum downward-view optical opacities
until we reach optical depth $\tau=1$ and $\tau=0.1$ to keep track the
evolution of this cloud's appearance as seen from a distance;
9) We return to step \#4 above and iterate another timestep $dt$, integrating
again the equations of motion.
We continue this algorithm until all gas is lost and all regolith has fallen to the
ground.

Finally, when all dust has fallen out, we calculate where the regolith ejecta
have been deposited.
Because we're representing each shell as a single point for the purposes of the
equations of motion calculations, we want to do more than simply plot the location
of each shell-particle on the ground to determine the deposition profile of ejected
regolith.
Thus we create a template function for the deposition of a ballistic explosion of a
single spherical shell of material.  By applying this template to each
shell-particle's final resting location, we better approximate the total deposition
of material from that shell.
We then sum all of the material from the $\approx 600,000$ shells to determine the
overall dust ejecta deposition profile.

There are obvious caveats to this calculation.
Undoubtably the release mechanism is more complex than that adopted here, but
this release mode is sufficiently simple to be modeled.
Secondly, the diffusion constant, and therefore the minimal flow rate, might be
overestimated due to the significant (but still largely unknown) decrease in
regolith porosity with increasing depth on the scale of meters (Carrier et
al.\ 1991), plus the liklihood that simulants used have larger particles and
greater porosity than typical regolith.
Lastly, the regolith depth of 15 meters might be an overestimate for some of
these regions, which are among the volcanically youngest and/or freshest
impacts on the lunar surface.
This exception does not apply to Plato and its active highland vicinity,
however, and Aristarchus is thickly covered with apparent pyroclastic deposits
which likely have different but unknown depths and diffusion characteristics.

\subsection{Results from Explosive Outgassing Model}

We find results for this ``minimal TLP'' numerical model of explosive
outgassing through the lunar regolith interesting in terms of the reported
properties of TLPs.
Figure 2 shows the evolution of the model explosion with time, as might be seen
from an observer above, in terms of the
optical depth $\tau = 1$ and $\tau = 0.1$ profiles of the model event, where
$\tau = 1$ is a rough measure of order unity changes in the appearance of the
surface features, whereas $\tau = 0.1$ is close to the threshold of the human
eye for changes in contrast, which is how many TLPs are detected (especially
the many without noticable color change).
In both cases the cloud at the particular $\tau$ threshold value expands
rapidly to a nearly fixed physical extent, and maintains this size until
sufficient dust has fallen out so as to prevent any part of the cloud from
obscuring the surface to this degree.

Easily-seen effects on features ($\tau = 1$) lasts for 50~s and extends over a
radius of 2 km, corresponding to 2 arcsec in diameter, resolvable by a typical
optical telescope but often only marginally so.
In contrast, the marginally detectable $\tau = 0.1$ feature extends over 14 km
diameter (7.5 arcsec), lasting for 90 s, but is easily resolved.
This model ``minimal TLP'' is an interesting match to the reported behavior of
non-instantaneous (not $\la 1$ s) TLPs: about 7\% of duration 90~s or less, and
half lasting under about 1300~s.
Certainly there should be selection biases suppressing reports of shorter
events.
Most TLP reports land in an envelope between about two minutes and one hour
duration, and this model event lands at the lower edge of this envelope.
Furthermore, most TLPs, particularly shorter ones, are marginally resolved
spatially, as would be the easily-detectable component of the model event.
This correspondence also seems interesting, given the simplicity of our model
and the state of ignorance regarding relevant parameters.

How might this dust cloud actually affect the appearance of the lunar surface?
First, the cloud should cast a shadow that will be even more observable than
simple surface obscuration, blocking the solar flux from an area comparable to
the $\tau = 1$ region and visible in many orientations.
Experiments with agitation of lunar regolith (Garlick et al.\ 1972b) show that
the reflectance of dust is nearly always increased under fluidization,
typically by about 20\% and often by about 50\% depending on the particular
orientation of the observer versus the light source and the cloud.
Similar results should be expected here for our simulated regolith cloud.
These increases in lunar surface brightness would be easily observable spread
over the many square kilometers indicated by our model.

Furthermore, because the sub-micron particle sizes dominate the outer regions
of the cloud, it seems reasonable to expect Mie-scattering effects in these
regions with both blue and red clouds expected from different Sun-Earth-Moon
orientations.
Figure 3 shows the typical fall-out time of dust particles as a function of
size.
Particles larger than $\sim$30 $\mu$m all fall out within the
first few seconds, whereas after a few tens of seconds, particles are
differentiated for radii capable of contributing to wavelength-dedendent
scattering.
Later in the event we should expect significant color shifts (albeit not
order-unity changes in flux ratios).

The larger dynamical effects in the explosion cloud change rapidly over the
event.
Half of the initially entrained gas is lost from the cloud in the first 3 s,
and 99\% is lost in the first 15~s.
Throughout the observable event, the remaining gas stays in good thermal
contact with the dust, which acts as an isothermal reservoir.
Gas escaping the outer portions of the dust cloud does so at nearly the sound
speed ($\sim 420$ m s$^{-1}$), and the outer shells of dust also contain
particles accelerated to similar velocities.
Gas escaping after about 3 s does so from the interior of the cloud in parcels
of gas with velocities decreasing roughly inversely with time.
One observable consequence of this is the expectation that much of the gas and
significant dust will be launched to altitudes up to about 50 km, where it may
be observed and might affect spacecraft in lunar orbit.

The longterm effects of the explosion are largely contained in the initial
explosion crater (nominally 14 m in radius), although exactly how the ejecta
ultimately settle in the crater is not handled by the model.
At larger radii the model is likely to be more reliable; Figure 4 shows how
much dust ejecta is deposited by the explosion as a function of radius.
Beyond the initial crater, the surface density of deposited material varies
roughly as $\mu \propto r^{-2.7}$, so it converges rapidly with distance.
Inside a radius of $\sim 300$ m, the covering factor of ejecta is greater than
unity; beyond this one expects coverage to be patchy.
This assumes that the crater explosion is symmetric and produces few ``rays.''
The explosion can change the reflectivity by excavating fresh material.
This would be evidenced by a $\sim$10\% drop in reflectance at wavelength
$\lambda \approx 950$~nm caused by surface Fe$^{2+}$ states in pyroxene and
similar minerals (Adams 1974, Charette et al.\ 1976).
Likewise there is an increase in reflectivity in bluer optical bands (Buratti
et al.\ 2000) over hundreds of nm.
Even though these photometric effects are compositionally dependent, we are
interested only in differential effects: gradients over small distances and
rapid changes in time.
The lifetime of even these effects at 300 m radius is short, however, due to
impact ``gardening'' turnover.
The half-life of the ejecta layer at 300 m radius is only of order 1000~y
(from Gault et al.\ 1974), and shorter at large radius (unless multiple
explosions accumulate multiple layers of ejecta).
At 30~m radius the half-life is of order 10$^6$~y.
From maturation studies of the 950~nm feature (Lucey et al.\ 2000, 1998), even
at 30~m, overturn predominates over optical maturation rates (over hundreds of
My).

The scale of outgassing in this model event, both in terms of gas release
($\ga$~1~tonne) and timescale ($\ga$~4~d), are consistent with the total
gas output and temporal granularity of outgassing seen in $^{40}$Ar, a dominant
lunar atmospheric component.
The fact that this model also recovers the scale of many features actually
reported for TLPs lends credence to the idea that outgassing and TLPs might be
related to each other causally in this way, as well as circumstantially via the
Rn$^{222}$ episodes and TLP geographical correlation (Paper I).

How often such an explosive puncturing of the regolith layer by outgassing
should
occur is unknown, due to the uncertainty in the magnitude and distribution of
endogenous gas flow to the surface, and to some degree how the regolith reacts
in detail to large gas flows propagating to the surface.
Also, a new crater caused by explosive outgassing will change the regolith
depth, its temperature structure, and eventually its diffusivity.
We will not attempt here to follow the next steps in the evolution of an
outgassing ``fumerole'' in this way, but are inspired to understand how
regolith, its temperature profile, and gas interact, as in the next section.
Furthermore, such outgassing might happen on much larger scales, or might over
time affect a larger area.
Indeed, such a hypothesis is offered for the scoured region of depleted
regolith forming the Ina D feature and may extend to other regions around
Imbrium (Schultz et al.\ 2006).

Our results here and in Paper I bear directly on the argument of Vondrak (1977)
that TLPs as outgassing events are inconsistent with SIDE episodic outgassing
results.
The detection limits from ALSEP sites {\it Apollo 12, 14} and {\it 15}
correspond to 16-71 tonne of gas per event at common TLP sites, particularly
Aristarchus.
(Vondrak states that given the uncertainties in gas transportation, these
levels are uncertain at the level of an order of magnitude.)~
Our ``minimal TLP event'' described above is 20-80 times less massive than
this, however, and still visible from Earth.
It seems implausible that a spectrum of such events would never exceed the SIDE
limit, but it is not so obvious such a large event would occur in the
seven-year ALSEP operations interval.
Also, this SIDE limit interpretation rests crucially on
Alphonsus (and Ross D) as prime TLP sites, both features which are rejected by
our robust geographical TLP sieve in Paper I.

\subsection{Coronal Discharge Effects}

Dust elutriation or particle segregation in a cloud agitated by a low-density
gas, occurring in this model, could potentially generate large electrostatic
voltages, perhaps relating to TLPs (Mills 1970).
Luminous discharges are generated in terrestrial volcanic dust clouds (Anderson
et al.\ 1965, Thomas et al.\ 2007).
Above we see dust particles remain suspended in a gas of
number density $n = 10^{10}$ to $10^{15}$ cm$^{-3}$ on scales of several tenths
of a km to several km, a plausible venue for large voltages.
In the heterogeneous lunar regolith, several predominant minerals with
differing particle size may segregate under gas flow suspension and
acceleration.
Assuming a typical particle size of $r=10~\mu$m, and typical work function
differences\footnote{
Predicting actual values of $\Delta W$ for particles of even well-defined
compositions is problematic due to surface effects such as
solar-wind/micrometeoritic weathering and exposed surface Fe$^{2+}$ states.
The following analysis suffices for two particles of different conducting
composition; a similar result arises via triboelectric interaction of two
different dielectrics although the details are less understood.
Disturbed dust is readily charged for long periods in the lunar surface
environment (Stubbs, Vondrak \& Farell 2005).}
of $\Delta W = 0.5$ eV, two particles exchange charge upon contact until
the equivalent of $\pm$0.25V is maintained, amounting to
$Q=CV=4\pi \epsilon_0 rV = 2.8 \times 10^{-16}$ coul = 1700 e$^-$.
When these particles separate to distance $d >> r$, their mutual
capacitance becomes $C_2=4\pi \epsilon_0 r^2/d$.
For $d = 100$~m, if particles retain $Q$, voltages increase by $10^7$ times!
Such voltages cannot be maintained.

Paschen's coronal discharge curve reaches minimum potential at 137V for Ar,
156V for He, for column densities $N$ of $3.2\times 10^{16}$ and $1.4\times
10^{17}$ cm$^{-2}$, respectively, and rises steeply for lesser column
densities (and roughly proportional to $N$ for larger $N$).\footnote{These
results depend somewhat the electrode's structure and composition.
In carefully controlled conditions, breakdown voltages in He as low as 20V
are achieved (Compton, Lilly \& Olmstead 1920).}
Similar optimal $N$ are found for molecules, with minimum
voltages a few times higher e.g., 420V at $1.8 \times 10^{16}$ cm$^{-2}$ for
CO$_2$, 414V for H$_2$S, 410V for CH$_4$, and $N$ for other
molecules $\sim 2.1 \times 10^{16}$ cm$^{-2}$.

The visual appearance of atomic emission in high voltage
discharge tubes is well know, with He glowing pink-orange (primarily at
4471.5\AA\ and 5875.7\AA: Reader \& Corliss 1980, Pearse \&
Gaydon 1963), and Ar glowing violet (from lines 4159-4880\AA).
If this applies to TLPs, the incidence of intense red emission in some TLP
reports (Cameron 1978) argues for another gas.\footnote{
The events and their colors reported can be counted
(although a variation in outcomes depending on how one
categorizes mixed colors and other factors).
Following Paper I, we divide the sample at year 1956 (the later composing 1/4
of the sample); after 1955/before 1956: red - 52/11, blue (or blue-green) - 0/7,
violet - 3/5, yellow - 0/4, red-yellow - 0/3, brown - 0/2, orange - 1/2.
Red is common (out of 894 reports total, not including
those by Bartlett, which are largely blue), however, there is a
statistically significant change after 1956, where about 1/5 of reports include
red color, many associated with Aristarchus, presumably due to a shift in
observer behavior rather than the physics of lunar events.
At least 20 nightside TLPs are reported, usually a bright and/or variable
spot, 8 at/near Aristarchus e.g.,
1824 May 1, near Aristarchus ``blinking light, 9th to 10th mag on dark side;
1881 February 3, near Aristarchus: ``very bright, like an 8th mag star,
pulsating;''
1789 May 29: ``flickering spot on east edge of Grimaldi,''  etc.
Earthshine lunar surface brightness is $\mu \approx 12$ to
13 (mag arcsec$^{-2}$) in V, compared to 3.4 at full Moon, so visible 
sources could be faint.}
Ne is not an endogenous gas.
Common candidate molecules appear white
or violet-white (CO$_2$, SO$_2$) or red (water vapor - primarily H$\alpha$,
which is produced in many hydrogen compounds; CH$_4$ - Balmer lines plus CH
bands at 390 and 431 nm).

The initial gas density at the surface from a minimal TLP is
$\sim 10^{18}$ cm$^{-3}$, so initially the optimal $N$ for coronal discharge
is on cm scales (versus the initial outburst over tens of meters).
As the TLP expands to 1 km radius, $n$ drops to $\sim 10^{13}$
cm$^{-3}$, so the optimal $N$ holds over the scale of the entire cloud,
likely the most favorable condition for coronal discharge.
If gas kinetic energy converts to luminescence with, for instance, 2\%
efficiency, at this density this amounts to $\sim 0.1$ J m$^{-3}$, or
100 J m$^{-2}$, compared to the reflected solar flux of 100 W m$^{-2}$,
capable of a visible color shift for several seconds.
Perhaps a minimal TLP could sustain a visible coronal discharge over much of
its $\sim 1$ min lifetime.
These should also be observable on the nightside surface, too, since
solar photoionization is seemingly unimportant in initiating the discharge,
and there are additional factors to consider.\footnote{
The timescale predicted for this effect depends largely on assuming
gas emission in a single burst.
Eruptions might occur along a non-point source, or from gas
backed up along the length of a channel, producing a prolonged eruption.
Additional complications might include interaction of agitated dust and/or
gas with the channel.
Such interactions can also generate much ionized gas under
moderate voltages e.g., via a mechanism similar to that in a
capillaritron (Perel et al.\ 1981, Bautsch et al.\ 1994).}

\section {Seepage of Gas Through the Regolith}

Referring back to scenarios (2) and (3) in \S1, the onset of fluidization
(Mills 1969) marks the division between these two regimes of seepage and
``bubbling'' and has been studied (Siegal \& Gold 1973, Schumm 1970).
Although laboratory test are made with coarser sieve particulates and much
thinner dust layers in 1 $g$ gravity, we can scale the gas pressure needed for
incipient fluidization by $g^{-2}$ and thickness $t^1$ to find the threshold
$P = 0.16$ atm (Siegal \& Gold 1973).
Correcting for less diffusive regolith, this pressure estimate is likely a
lower limit.
Below this pressure simple gas percolation likely predominates.

What processes occur during ``simple'' percolation?
Were it not for phase changes of venting gas within the regolith, the
composition of the gas might be a weak consideration in this paper (except
for perhaps the molecular/atomic mass), and temperature would likely only
affect seepage as $T^{1/2}$ in the diffusivity.
Water plays a special role in this study (separate from concerns regarding
resource exploitation or astrobiology), in that it is the only common substance
encountering its triple point temperature in passing through the regolith, at
least in many locations.
In this case water might not contribute to overpressure underneath the regolith
leading to explosive outgassing.
This would also imply that even relatively small volatile flows containing
water would tend to freeze in place and remain until after the flow stops.
For water this occurs at 0.01$^\circ$C, corresponding to 0.006 atm in pressure
(the pressure dropping by a factor of 10 every $\sim$25$^\circ$.)~

Effectively, water is the only relevant substance to behave in this fashion.
The next most common substances may be large hydrocarbons such as nonane or
benzene, obviously not likely abundant endogenous effluents from the interior.
Also H$_2$SO$_4$ reaches its triple point, but changes radically with even
modest concentrations of water.
A similar statement can be made about HNO$_3$, not a likely outgassing
constituent.
These will not behave as their pure state, either; this leaves only H$_2$O.
Water (and sulfur) has been found in significant
concentration in volcanic glasses from the deep lunar interior (Saal et
al.\ 2008, Friedman et al.\ 2009), and has been liberated in large quantities in
past volcanic eruptions.
The measured quantities of tens of ppm imply juvenile concentrations of
hundreds of ppm.

From the heat flow measurements at the {\it Apollo 15} and {\it 17} lunar
surface experiment (ALSEP) sites (Langseth \& Keihm 1977), we know that just
below the surface, the stable regolith temperature is in the range of 247-253K
(dependent on latitude, of course), with gradients (below 1-2 m) of 1.2-1.8 deg
m$^{-1}$, which extrapolates to $0^\circ$C at $13-16$ m depths subsurface.
With the exception of the outermost few centimeters, the entire regolith is
below the triple point temperature and is too deep to be affected
significantly by variations in heating over monthly timescales.
This is an interesting depth, since in many areas the regolith is not quite
this deep, as small as under a few meters near Lichtenberg (Schultz \& Spudis
1983) and at the Surveyor 1 site near Flamsteed (Shoemaker \& Morris 1970) to
depths at Apollo sites (summarized by McKay et al.\ 1991) near the $0^\circ$C
depths calculated above, up to probably 20 m or more in the highlands, and
40~m deep north of the South Pole-Aitken Basin (Bart \& Melosh 2005).
Presumably, the fractured megaregolith supporting the regolith likely does not
contain as many small particles useful for retaining water ice, as we detail
below, but it may accumulate ice temporarily.
Recent heat flow analyses (Saito et al.\ 2007) account for longer
timescale fluctuations placing the $0^\circ$C depth twice as far
subsurface, increasing the lifetime of retained volatiles against
sublimation accordingly; for now we proceed with a more conventional,
shorter-lived analysis.

The escape of water and other volatiles into the vacuum is regulated by the
state of the regolith and is presumably largely diffusive.
We assume the Knudsen flow regime (low-density, non-collisional gas).
Of special importance is the measured abundance of small dust grains
in the upper levels of the regolith, which perhaps pertains to depths $\sim
15$~m (where bulk density is probably higher: Carrier et al.\ 1991).
Assuming that particle distributions are self-similar in size distribution
(constant porosity), for random-walk diffusion out of a volume element $dV$,
the diffusion time step presumably scales with the particle size $a$, so the
diffusion time $t \propto a^{-1}$.
For particles of the same density, therefore, one should compute the diffusion
time by taking a $a^{-1}$-weighted average of particle sizes counted by mass,
$\langle a \rangle$.
This same moment of the distribution is relevant in \S 2.

Published size distributions measured to sufficiently small sizes include again
McKay et al.~(1974) with $\langle a \rangle = 24$ $\mu$m, and supplemented on
smaller sizes with {\it Apollo 11} sample 10018 (Basu \& Molinaroli 2001),
which reduces the average to about 20 $\mu$m.
This is an overestimate because a large fraction (34-63\%) are agglutinates,
which are groupings of much smaller particles.
Many agglutinates have large
effective areas e.g., $e = A/4 \pi r^2$, with values of a few up to 8.
(Here $r$ is a mean radius from the center of mass to a surface element.)~
To a gas particle, the sub-particle size is more relevant than the agglutinate
size, so the effective particle size of the entire sample might be much
smaller, conceivably by a factor of a few.

We compare this to experimental simulations, a reasonably close analogy being
the sublimation of a slab of ice buried up to 0.2 m below a medium of simulant
JSC Mars-1 (Allen et al.\ 1998) operating at $\sim 263^\circ$K and 7 mbar
(Chevrier et al.~2007), close to lunar regolith conditions.
This corresponds to the lifetime of 800~y for a 1 m thick ice layer covered by
1~m of regolith.
The porosity of JSC Mars-1 is 44-54\%, depending on compactification whereas
lunar soil has $\sim$49\% at the surface, perhaps 40\% at a depth of 60 cm,
and slightly lower at large depths (Carrier et al.\ 1991).
Lunar soil is somewhat less diffusive by solely this measure.
The mean size $\langle a \rangle$ of JSC Mars-1 is 93 $\mu$m, $\ga$10 times
larger than that for {\it Apollo 17} and {\it 11} regolith, accounting for
agglutinates, so the sublimation timescale for regolith material is, very
approximately, $\ga$10 ky (perhaps up to $\sim$30 ky).
Other simulants are more analogous to lunar regolith, so future
experiments might be more closely relevant.

Converting a loss rate for 1 m below the surface to 15~m involves the
depth ratio $R$.
Farmer (1976) predicts an evaporation rate scaling as $R^{-1}$ (as opposed to
the no-overburden analysis: Ingersoll 1970).
Experiments with varying depths of simulated regolith (Chevrier et al.\ 2007)
show that the variation in lifetime indeed goes roughly as $R^1$, implying a
1~m ice slab lifetime at 15~m on the order of $10^5$ to $3 \times 10^5$ y.
The vapor pressure for water ice drops a factor of 10 in passing from
$0^\circ$C to current temperatures of about $-23^\circ$C just below the surface
(also the naked-ice sublimation rate: Andreas 2007), which would indicate that
$\sim$90\% of water vapor tends to stick in overlying layers (without affecting
the lifetime of the original layer, coincidentally).
This begs the question of the preferred depth for an ice layer to form.
The regolith porosity decreases significantly between zero and 1~m depth
(Carrier et al.\ 1991) which argues weakly for preferred formation at greater
depth.
At 30~m depth or more, the force of overburden tends to close off porosity.

The current best limit on water abundance is from the sunrise terminator
abundances from LACE, which produces a number ratio of H$_2$O/$^{40}$Ar with a
central value of 0.014 (with $2\sigma$ limits of 0-0.04).
This potentially indicates
an actual H$_2$O/$^{40}$Ar outgassing rate ratio up to 5 times higher (Hoffman
\& Hodges 1975).
Adopting the SIDE rate of 7 g s$^{-1}$ in the $\sim$20-44 AMU mass range, and
assuming most of this is $^{40}$Ar (Vondrak, Freeman \& Lindeman 1974: given
the much lower solar wind contributions of other species in this range), this
translates to 0.1 g s$^{-1}$ of water (perhaps up to 0.5 g s$^{-1}$ or
15 tonne y$^{-1}$), in which case most of the gas must be ionized.
The disagreement between SIDE and LACE is a major source of uncertainty
(perhaps due to the neutral/ionized component ambiguity).

We discuss below that at earlier times the subsurface temperature was likely
lower, but let us consider now the situation in which a source arises into
pre-established regolith in recent times.
We assume a planar diffusion geometry, again.
In this case, we take spatial gradients over 15~m and scale the JSC Mars-1
diffusivity of 1.7~cm$^2$ s$^{-1}$ to 0.17~cm$^2$ s$^{-1}$ for lunar regolith.
Since the triple-point pressure corresponds to number density $n = 2.4 \times
10^{17}$ cm$^{-3}$, the areal particle flux density is $j = 2.7 \times 10^{13}$
s$^{-1}$ cm$^{-2}$.
For a large outgassing site, with the same water fraction of water indicated by
LACE e.g., total outgassing of 7 g s$^{-1}$ including 0.1 g s$^{-1}$ of water,
this rate can maintain a total area of 0.012 km$^2$ at the triple-point
pressure i.e., a 125~m diameter patch.
This is much larger than the 15~m regolith depth, bearing out our assumed
geometry.
If this ice patch were 1~m thick, for example, the ice would need to be
replenished every 4000~y.

Of course, this is a simple model and many complications could enter.
We consider briefly the effects of latitude, change in lunar
surface temperature over geological time, and the effects of aqueous chemistry
on the regolith.

\subsection{Latitude Effects}

The temperature just below the surface is legislated by the time-averaged
energy flux in sunlight, so it scales according to the Stefan-Boltzmann law
from the temperature at the equator ($\phi = 0$) according to $T (\phi) = T(0)
cos^{1/4} (\phi)$.
This predicts a 6K temperature drop from the equator (at about 252K) to the
latitude of the Aristarchus Plateau ($\phi \approx 26^\circ$) or the most polar
subsurface temperature measurement by {\it Apollo 15}, a drop to 224K at Plato
($\phi = 51^\circ.6$), $\la$200K for the coldest 10\% of the lunar surface
($\phi > 65^\circ$) and $\la$150K for the coldest 1\%
($\phi > 82^\circ$).\footnote{The most polar maria edge might border
Frigoris at $\phi \approx 65^\circ$, but there are flooded craters
much higher.}
These translate into a regolith depth at the water triple-point of
$\sim$4, 18, 33 or 65 m deeper than at the equator, respectively, probably
deeper in the latter cases than the actual regolith layer.
Permanently shadowed cold traps, covering perhaps 0.1\% of the surface, have
temperatures $\la$60K (e.g., Adorjan 1970, Hodges 1980).
(Note that the lunar South Pole is a minor TLP site responsible for about 1\%
of robust report counts.)~
Since even at the equator the H$_2$O triple point temperature occurs $\sim$13~m
below the surface, at increasing latitude this zone quickly moves into the
megaregolith where the diffusivity is largely
unknown but presumably higher (neglecting the decrease in porosity due to
compression by overburden).
To study this, we assume a low diffusivity regolith layer 15~m deep
overlying a high diffusivity layer which may contain channels
directing gas quickly upward (although perhaps not so easily
horizontally).

The diffusivity of the regolith near 0$^\circ$C is dominated by elastic
reflection from mineral surfaces, without sticking, whereas at lower
temperatures H$_2$O molecules stick during most collisions (Haynes, Tro \&
George 1992).
This is especially the case if the surfaces are coated with at least a few
molecular layers of H$_2$O molecules, of negligible mass.
The sticking behavior of H$_2$O molecules on water ice has been studied over
most of the temperature range relevant here (Washburn et al.\ 2003); but does
depend somewhat on whether the ice is crystalline or amorphous (Speedy et
al.\ 1996).
In contrast the sticking behavior of H$_2$O molecules on lunar minerals is much
less well known.

The lunar simulant diffusivity value above corresponds to a mean free path
time of $\sim$1 $\mu$s for H$_2$O molecules near 0$^\circ$C.
In contrast the timescale for H$_2$O molecules sticking on ice is
(from Schorghofer \& Taylor [2007] and references therein): $\tau = \theta /
( \alpha P_v / \sqrt { 2 \pi k T \mu } ),$
where $\theta$ is the areal density of H$_2$O molecules on ice $\theta =
( \rho / \mu )^{2/3} \approx 10^{15}$ cm$^{-2}$ for density $\rho$, and
molecular mass $\mu$.
The sticking fraction $\alpha$ varies from about 70\% to 100\% for $T = 40$K to
120K.
The equilibrium vapor pressure is given by $P_v = p_t ~ exp [ - Q ({1 \over T}
- {1 \over T_t}) / k ]$ where $p_t$ and $T_t$ are the triple point pressure and
temperature, respectively, and sublimation enthalpy $Q = 51.058$ kJ/mole.
This expression and laboratory measurements imply a sticking timescale $\tau
\approx 1~ \mu$s at $T = 260$K, 1 ms at 200K, 1 s at 165K, 1 hr at 134K, and
1 yr at 113K.
The sticking timescale quickly and drastically overwhelms the kinetic
timescale at lower temperatures.

This molecular behavior has a strong effect on the size of the ice patch
maintained by the example source considered above.
Simply scaling by the time between molecular collisions, corresponding to a
125~m diameter ice patch at $\phi = 0$, we find at the base of the regolith a
160~m patch at $\phi = 26^\circ$ (Aristarchus Plateau),
580~m at $\phi = 51^\circ .6$ (Plato),
2.3~km at $\phi = 65^\circ$ (10\% polar cap),
and an essentially divergent value, 522~km at $\phi =82^\circ$ (1\% polar cap).
If in fact the regolith layer is much deeper than suspected, the added depth of
low diffusivity dust significantly increases the patch area: 170~m at $\phi =
26^\circ$, 830~m at $\phi = 51^\circ.6$, and 4~km at $\phi = 65^\circ$.
Figure 5 presents graphically how the growth of the ice patch varies with
latitude, plus also the effects of flow rate and the assumed regolith depth.

\subsection{Longterm Evolution}

Most portions of the lunar surface have been been largely geologically inactive
during the past 3 Gy or more (with some of the notable exceptions listed above).
During this time several important modifications of the scenario above are
relevant.

The current heat flow from the lunar interior, $j_{int} \approx 0.03$ W
m$^{-2}$ (Langseth et al.\ 1972, 1973), is only a $2 \times 10^{-5}$
part of the solar constant, so it affects the temperature near the lunar
surface at the level of only $\sim 1.3$ millidegree.
There were times in the past, however, when interior heating likely pushed
the temperature near the surface over $0^\circ$C.

A zero-degree zone near the maria presumably could not form
until $\sim$3 Gy ago, probably sufficient for the Moon globally (see
Spohn et al.\ 2001).
After this the $0^\circ$C depth receded into the regolith, while the
regolith layer was also growing.
Simultaneously, the average surface temperature was cooler by
$\sim15$ degree due to standard solar evolution (Gough 1981 -- perhaps
$17^\circ$ lower in the highlands at 4 Gy ago).
Since the the thickness of regolith after 3 Gy ago grows at only about
1 m per Gy (Quaide \& Oberbeck 1975), within the maria the $0^\circ$C depth
sinks into bedrock/fractured zone.
Whatever interaction and modification might be involved between the regolith
and volatiles will proceed inwards, leaving previous epochs' effects between
the surface vacuum and the $0^\circ$C layer now at $\sim 15$~m.

\subsection {Chemical Interactions with Regolith}

Another issue to consider
is possible regolithic chemical reactions with outgassing
volatiles, especially over prolonged geological timescales.
The key issue is the possible presence of water vapor, and perhaps SO$_2$.
There is little experimental work on the aqueous chemistry of lunar regolith
(which will vary due to spatial inhomogeneity).
Dissolution of lunar fines by water vapor is greatly accelerated in the absence
of other gases such as O$_2$ and N$_2$ (Gammage \& Holmes 1975) and appears to
proceed by etching the numerous damage tracks from solar-wind particles.
This process acts in a way to spread material from existing grains without
reducing their size (which would otherwise tend to increase porosity).
Liquid water is more effective than vapor, not surprisingly, and ice tends to
establish a pseudo-liquid layer on its surface.
This is separate from any discussion of water retention on hydrated minerals
surfaces robust to temperatures above 500$^\circ$C (Cocks et al.\ 2002 \& op
cit.).

This is a complex chemical system that will probably not be understood without
simulation experiments.
The major constituents are presumably silicates, which will migrate in
solution only over geologic time.
(On Earth, consider relative timescales of order
30 My typical migration times for quartz,
700 ky for orthoclase feldspar, KAlSi$_3$O$_8$ and
80 ky for anorthite, CaAl$_2$Si$_2$O$_8$: Brantley 2004.)~
One might also expect the production of Ca(OH)$_2$, plus perhaps Mg(OH)$_2$
and Fe(OH)$_2$.
It is not clear that Fe(OH)$_2$ would oxidize to more insoluble FeO(OH), but
any free electrons would tend to encourage this.
It seems that the result would be generally alkaline.
Since feldspar appears to be a major component in some outgassing regions
e.g., Aristarchus (McEwen et al.\ 1994), one should also anticipate the
production of clays.
This is not accounting for water reactions with other volatiles e.g., ammonia,
which has been observed as a trace gas (Hoffman \& Hodges 1975) perhaps in
part endogenous to the Moon, and which near $0^\circ$C can dissolve in water
at nearly unit mass ratio (also to make an alkaline solution).
Carbon dioxide is a possible volatile constituent, and along with water can
metamorphose olivine/pyroxene into Mg$_3$Si$_4$O$_{10}$(OH)$_2$ i.e. talc,
albeit slowly under these conditions; in general the presence of CO$_2$ and
thereby H$_2$CO$_3$ opens a wide range of possible reactions into carbonates.
Likewise the presence of sulfur (or SO$_2$) opens many possibilities e.g.,
CaSO$_4 \cdot$2H$_2$O (gypsum), etc.
Since we do not know the composition of outgassing volatiles in detail, we
will probably need to inform simulation experiments with further remote
sensing or in situ measurements.

The mechanical properties of this processed regolith are difficult to predict.
Some possible products have very low hardness and not high ductility.
Some of these products expand but will likely fill the interstitial volume
with material, which will raise its density and make it more homogeneous.
Regolith is already ideal in having a nearly power-law particle distribution
with many small particles.
It seems likely that any such void-filling will sharply reduce diffusivity.
The volatiles actually discovered in volcanic glasses from the deep interior
(Saal et al.\ 2008, Friedman et al.\ 2009) include primarily H$_2$O and SO$_2$
but not CO$_2$ or CO.
With the addition of water, regolithic mineral combinations tend to be
cement-like, and experiments with anorthositic lunar chemical simulants have
produced high quality cement without addition of other substances, except
SiO$_2$ (Horiguchi et al.\ 1996, 1998).
Whether this happens $in$ $situ$ depends upon whether over geological time
(CaO)$_3$SiO$_2$ or other Ca can act as a binder without heating to sintering
temperatures.
The possible production of gypsum due to the high
concentrations of sulfur would add to this cement-like quality.
The extent to which ordinary mixes such as portland cement lose water into the
vacuum depend on their content of expansive admixture (Kanamori 1995).
Portland cement mixes show little evidence of loss of compressional strength
in a vacuum (Cullingford \& Keller 1992).

We need to think in terms of possibly cemented slabs in some vicinities, and
need to consider the effects of cracks or impacts into this concrete medium.
This is probably not a dominant process, since the overturn timescale to
depths even as shallow as 1~m is more than 1~Gy (Gault et al.\ 1974, Quaide \&
Oberbeck 1975), whereas we discuss processes at $\sim$15~m or more.
Craters 75~m in diameter will permanently excavate to a 15~m depth (e.g.,
Collins 2001, and ignoring the effects of fractures and breccia formation), and
are formed at a rate of about 1 Gy$^{-1}$ km$^{-2}$ (extending Neukum et
al.\ [2001] with a Shoemaker number/size power-law index 2.9).
This will affect some of the areal scales discussed above, but not all.
We speculate that vapor or solution flow might tend to deliver ice and/or
solute to these areas and eventually act to isolate the system from the vacuum.
Finally, we note above that over geological timescales this ice layer will
tend to sink slowly into the regolith, at a rate of order 1~m Gy$^{-1}$,
setting up a situation where any relatively impermeable concrete zone will
tend to isolate volatiles from the vacuum.
In this case volatile leakage will tend to be reduced to a peripheral region
around the ice patch.
Assuming that volatiles leak out through the entire 15~m thickness of regolith
at the patch boundary, the 125~m diameter patch area for $\phi \approx 0$ from
above corresponds to a peripheral zone expected from a 520~m diameter patch.
Thus any such concrete overburden will encourage growth of small patches,
and will do so even more for larger ones (assuming $\la$15~m regolith depth).

How much water might reasonably be expected to outgas at these sites?
The earliest analyses of Apollo samples argued for extreme scarcity of
water and other volatiles (Anders 1970, Charles, Hewitt \& Wones 1971,
Epstein \& Taylor 1972).
On the Earth, water is the predominant juvenile outgassing component (Gerlach
\& Graeber 1985, Rubey 1964), whereas even the highest water concentrations
discussed below (Saal et al.\ 2008) imply values an order of magnitude
smaller.
On the Moon, water content is drastically smaller, with a current atmospheric
water content much less than what would affect hydration in lunar minerals
(Mukherkjee \& Siscoe 1973), although some lunar minerals seem to involve water
in their formation environments (Agrell et al.\ 1972, Williams \& Gibson 1972,
Gibson \& Moore 1973, and perhaps Akhmanova et al.\ 1978).
The origin of the water in volcanic glasses (Saal et al.\ 2008, Friedman et
al.\ 2009) is still poorly understood but implies internal concentrations that
at first look seems in contradiction with earlier limits e.g., Anders 1970.
For much different lunar minerals, high concentration is implied for water
(McCubbin et al.\ 2007) as well as other volatiles (Kr\"ahenb\"uhl et
al.\ 1973).

It is not a goal of this paper to explain detected water in lunar samples, but
its origin at great depth is salient here.
As a point of reference, Hodges and Hoffman (1975) show that the $^{40}$Ar in
the lunar atmosphere derives from deep in the interior, of order 100~km or
more.
They hypothesize that the gas could just as easily derive from the
asthenosphere, 1000~km deep or more (see also Hodges 1977).
The picritic glasses analyzed by Saal, Friedman, et al.\ derive from depths of
$\sim$300-400 km or greater (Elkins-Tanton et al.\ 2003, Shearer, Layne \& 
Papike 1994).
O$_2$ fugacity measurements e.g., Sato 1979, are based on glasses from equal or
lesser depths.
Water originating from below the magma ocean might provide one
explanation (Saal et al.\ 2008), as might inhomogeneity over the lunar surface.
Differentiation might not have cleared volatiles from the deep interior
despite its depletion partially into the mantle.
One might also consider that geographical variation between terranes e.g.,
KREEP (K-Rare Earth Element-P) or not, might be important.

In the Moon's formation temperatures of proto-Earth and progenitor impactor
material in simulations grow to thousands of Kelvins, sufficient to drive off
the great majority of all volatiles, but these are not necessarily the only
masses in the system.
Either body might have been
orbited by satellites containing appreciable volatiles,
which would likely not be heated to a great degree and which would have had a
significant probability of being incorporated into the final moon.
Furthermore, there is recent discussion of significant water being delivered to
Earth/Moon distances from the Sun in the minerals themselves (Lunine et
al.\ 2007, Drake \& Stimpfl 2007), and these remaining mineral-bound even at
high temperatures up to 1000K (Stimpfl et al.\ 2007).
The volume of surface water on Earth is at least $1.4 \times 10^9$ km$^3$, so
even if the specific abundance of lunar water is depleted to $10^{-6}$
terrestrial, one should still expect over $10^{10}$ tonnes endogenous to the
Moon, and it is unclear that later differentiation would eliminate this.
This residual quantity of water would be more than sufficient to concern us
with the regolith seepage processes outlined above.

For carbon compounds, models of the gas filling basaltic vesicles
(Sato 1976; also O'Hara 2000, Wilson \& Head 2003, Taylor 1975) predict
CO, COS, and perhaps CO$_2$ as major components.
Negligible CO$_2$ is found in fire-fountain glasses originating from
the deep interior (Saal et al.\ 2008); this should be considered in light of CO
on the Moon (and CO$_2$ on Earth) forming the likely predominant gas driving
the eruption (Rutherford \& Papale 2009).

We suspect that water outgassing was likely higher in the past than it is now.
Furthermore, no site of activity traced by $^{222}$Rn or by robust TLP counts
(Paper I) has been sampled.
(The sample return closest to Aristarchus, {\it Apollo 12}, is 1100~km
away.)\footnote{
The only persistent TLP site of some statistical significance that corresponds
to a sample return (170 g from {\it Luna 24}) is Mare Crisium, ambiguous in
several ways:
1) the robust TLP report count for Crisium is only zero
to 6, depending on the robustness filter employed (Paper I);
2) the nature of this robust count is problematic given the extended nature of
Crisium as a feature, and
3) the 2-meter core sample returned from Crisium by {\it Luna 24} is one case
where the presence of significant water may be indicated (Akhmanova et
al.\ 1978).}
From our discussion above the behavior of outgassing sites near the poles
versus near the equator might differ greatly, with volatile retention near the
poles being long-term and perhaps making the processing of volatiles much more
subterranean and covert.
(Note that the lunar South Pole is a minor TLP site responsible for about 1\%
of robust report counts, as per Paper I.)~

With these uncertainties we feel unable to predict exactly how or
where particular evidence of lunar surface outgassing might be found,
although the results from above offer specific and varied signals
that might be targeted at the lunar surface.
For this reason we turn attention to how such effects
might be detected realistically from the Earth, lunar orbit, and near
the Moon's surface, and we suggest strategies not only for how these
might be tested but also how targeted observations might economically provide
vital information about the nature of lunar outgassing.
We consider the impact of recent hydration detections in \S 5.

We appreciate the controversial nature of suggesting small but
significant patches of subsurface water ice, given the history of the topic.
We take care to avoid ``cargo cult science'' - selection of data and
interpretion to produce dramatic but subjectively biased conclusions that do
not withstand further objective scrutiny (Feynman 1974).
Despite the advances made primarily by Apollo-era research, we are still
skirting the frontiers of ignorance.
We are operating in many cases in a regime where interesting observations have
been made but the parameters e.g., the endogenous lunar molecular production
(water vapor or otherwise), required to evaluate alternative models and
interpretations are sufficiently uncertain to frustrate immediate progress.
Below we offer several straightforward and prompt tests of our conclusions and
hypotheses which offer prospects of settling many of these issues.

\section{Observational and Experimental Techniques}

TLPs are rare and short-lived, which hampers their study.
We advance supplanting the current anecdotal catalog with data with {\it a
priori} explicit, calculable selection effects.
This might seem daunting; Paper I used in essence all known
reports from lunar visual observers since the telescope's invention!
With modern imaging and computing, it is tractable.

Another problem clear above is the variety of ways in which outgassing can
interact with the regolith.
In cases of slow seepage, gases may long delay their escape from the regolith.
If the gases are volcanic, they might interact along the way, and water vapor
might trap it and other gases in the regolith.
These factors bear on designing future investigations.

We can make significant headway exploiting more modern technology.
Table 1 lists the many methods detailed in this section.
There has been no areal-encompassing, digital image monitoring of the Near Side
with appreciable time coverage using modern software techniques to isolate
transients.
Numerous particle detection methods are promising.
The relevant experiments on Apollo were limited in duration, a week or less, or
5-8 years in the case of ALSEP.
Furthermore {\it Clementine} and the relevant portion of {\it Lunar Prospector}
were also relatively short. 
These limitations serve as background to the following discussions.
In this section we provide a potential roadmap to detailed study of outgassing.

\vskip -0.5in

\subsection{Optical/Infrared Remote Sensing}

Optical imaging advances several goals.
Transient monitoring recreates how TLPs were originally reported.
Not yet knowing TLP emission spectra, our bandpass should span the visual,
400-700 nm.
After an event, surface morphology/photometry changes might
persist, betrayed by 0.95 and 1.9 $\mu$m surface Fe$^{2+}$ bands and increased
blue reflectivity (\S 2.2).
Hydration is manifest in the infrared.
Asteroidal regolith 3~$\mu$m hydration signals are common (Lebofsky et
al.\ 1981, Rivkin et al.\ 1995, 2002, Volquardsen et al.\ 2004), and stronger
than those at 700 nm (Vilas et al.~1999) seen in lunar polar regions.
Absorption near 3 $\mu$m appears in lunar samples exposed to terrestrial
atmosphere for a few years (Markov et al.~1980, Pieters et al.\ 2005) but not
immediately (Akhmanova et al.~1972), disappearing within a few days in a dry
environment.
Further sample experiments are needed.

\vskip -0.5in

\subsubsection {Earth-Based Imaging}

Earth-based monitoring favors the Near Side, as do TLP-correlated effects:
$^{222}$Rn outgassing (all four events on nearside, plus most $^{210}$Po
residual) and mare edges.
The best, consistent resolution comes from the $Hubble$ $Space$ $Telescope$
with $0.07-0.1$ arcsec FWHM ($\sim$150~m) but with large overhead times.
Competing high-resolution imaging from ``Lucky Exposures'' (LE, also
``Lucky Imaging'') exploits occasionally superlative imaging within
a series of rapid exposures (Fried 1978, Tubbs 2003).
Amateur setups achieve excellent LE results, and the Cambridge group
(Law, Mackay \& Baldwin 2006) attains diffraction-limited imaging on a
2.5-meter telescope, $\sim$200-300~m FWHM.
Only $<$1\% of observing time survives image selection, but for the Moon this
requires little time.
LE resolution is limited to a seeing isoplanatic patch, $\sim$1000 arcsec$^2$,
3000 times smaller than the Moon.
Likewise, $HST$'s Wide Field Camera 3, covering 3000 arcsec$^2$, cannot
practically survey the Near Side.

\vskip -0.1in

High resolution imaging can monitor small areas over time or in one-shot
applications compared with other sources i.e., lunar imaging missions.
LE or $HST$ match the resolution of global maps from {\it Lunar
Reconnaissance Orbiter} Camera's (LROC) Wide-Angle Camera (Robinson et
al.\ 2005), and $Clementine$/UVIS, over 0.3-1 $\mu$m.
$Kaguya$'s Multiband Imager (Ohtake et al.\ 2007) has 40-m 2-pixel resolution.
LROC's Narrow Angle Camera has 2~m resolution in one band, targeted.
{\it Chang'e-1}/CCD (Yue et al.\ 2007) might also aid ``before/after''
sequences.
Lunar Orbiter images, resolving to $\sim$1~m, form
excellent ``before'' data for many sites, for morphological changes
e.g., cores of explosive events over 40 years.

The prime technique for detecting changes between epochs of similar images
is image subtraction, standard in studying supernovae, microlensing and
variable stars.
This produces photon Poisson noise-limited performance (Tomaney
\& Crotts 1996) and is well-matched to CCD or CMOS imagers, which at 1-2~arcsec
FWHM resolution cover the Moon with 10-20 Mpixels, readily available.
One needs $\ga$2 pixels FWHM, otherwise non-Poisson residuals dominate.
Our group has automated TLP monitors on the summit of Cerro Tololo, Chile and
at Rutherfurd Observatory in New York that produce regular lunar imaging
(Crotts et al.\ 2009), often simultaneously.
Each cover the Near Side at 0.6~arcsec/pixel with images processed in 10s.
This is sufficient to time-sample nearly all reported TLPs (see Paper I) and
produce residual images free of systematic errors at Poisson levels (Figure 6).
\footnote{
Multiple monitors can image TLPs in different bands or polarizations.
TLP polarimetric anomalies (Dollfus 2000) occur on uncertain timescales
(0.01-1~d).
Other polarimetric transients (Dzhapiashvili \& Ksanfomaliti 1962, Lipsky
\& Pospergelis 1966) are less constrained temporally.
These are likely scattering linear polarizations;
one can align one monitor's polarizer E-vector parallel to the
Sun-Moon direction on the sky, and a second perpendicular.
Three or four can construct linear Stokes parameters conventionally.
Also we plan a parallel video channel which dumps
high-speed sequences to disk given a TLP alert trigger.}

Imaging monitors open several possibilities for TLP studies, with
extensive, objective records of changes in lunar
appearance, at sensitivity levels $\sim$10 times better than the human eye.
An automated system can distinguish contrast changes of 1\% or
better, whereas the human eye is limited to $\ga$10\%.
We will measure the frequency of TLPs soon enough; Paper I indicates perhaps
one TLP per month visible to a human observing at full duty cycle.
TLP monitors open new potential to alert other observers, triggering LE imaging
of an active area, or spectroscopy of non-thermal processes and the gas
associated with TLPs.

\subsubsection {Ground-Based Spectroscopic/Hyperspectral Observations}

Spatially resolved spectroscopy can 1) elucidate TLP physics,
including identification of gas released, or 2) probe quasi-permanent changes
in TLP sites.
We must find changes in a four-dimensional dataset: two spatial
dimensions, wavelength, and time, too much to monitor for transients.
Fortunately, TLP monitoring can alert to an event in under 1000s, and a larger
telescope with a spectrograph can observe the target (within $\sim$300s).

Whereas ``hyperspectral'' imaging usually refers to resolving power
$R = \lambda / \Delta \lambda \approx 50-100$, where $\Delta \lambda$ is the
FWHM resolution, TLP emission might be much narrower, thereby diluted at low
resolution.
For line emission, rejecting photons beyond the line profile yields contrasts
up to 10$^4$ times better than the human eye using a telescope.
IR hydration band near 3.4~$\mu$m have substructure over $\sim$20~nm, requiring
$R \ga 300$, compared to the IR SpeX on the NASA Infrared Telescope Facility
with $R\la 2000$.
The 950~nm and 1.9 $\mu$m pyroxene bands show compositional shifts (Hazen, Bell
\& Mao 1978) seen at $R \approx 100$.
Differentiating pyroxenes from Fe-bearing glass (Farr et al.~1980) requires
$R \approx 50$.
Observations involve scanning across the lunar face with a long slit
spectrograph (Figure 7a).
Since lunar surface spectral reflectance is homogenized by
impact mixing, $>$99\% of the light in such a spectrum is ``subtracted away''
by imposing this average spectrum and looking for deviations (Figure 7b).
The data cube can be sliced in any wavelength to construct maps of lunar
features in various bands.
Figure 8 shows that surface features are reconstructed in detail and
fidelity.

What narrow lines might we search for?
The emission measure of gas in our model excited by solar radiation is
undetectable except for the first few seconds.
Coronal discharge offers a caveat.
Reddish discharge may indicate H$\alpha$ from dissociation of many
possible molecules.\footnote{Kozyrev (1963) reported transient H$_2$ emission
from Aristarchus (absent Balmer lines), and transient C$_2$ Swan bands (Kozyrev
1958).
We do not advance a model to explain these observations.}
Rather than relying on H$\alpha$ plus faint optical
lines/bands to distinguish molecules, note that near-IR
vibrational/rotational bands are brighter and more discriminatory.

\subsubsection {Surface and Subsurface Radar}

As in \S 3, internal water vapor might have produced ice in the
regolith $\la$15m subsurface, a venue for ground-penetrating radar, from
lunar orbit.
While epithermal neutrons and gamma radiation can detect hydrogen, they cannot
penetrate $\ga$1 m.
Near the poles or subject to chemical modification (\S 3.3), ice might range
closer to the surface.
Past and current lunar radar include {\it Apollo 17}'s Lunar Sounder Experiment
(LSE) (Brown 1972, Porcello 1974) at 5, 16 and 260 MHz, $Kaguya$'s Lunar Radar
Sounder (Ono \& Oya 2000) at 5 MHz (optionally, 1 MHz or 15 MHz), $LRO$'s
Mini-RF (Mini Radio-Frequency Technology Demonstration) at 3 and $\sim$10 GHz,
and Mini-SAR on {\it Chandrayaan-1} at 3 GHz (Bussey et al.\ 2006).
Shorter wavelength radar could map possible changes in surface features in
explosive outgassing, over tens of meters, in before/after radar sequences
meshed with optical monitoring e.g., with {\it LRO} Mini-RF (Chin et al.~2007).
For regolith and shallow bedrock, we need $\sim$100-300 MHz;
LSE operated only a few orbits and near the equator.
Near Side maps at $\sim$1~km resolution at 430~GHz (Campbell et al.~2007) could
improve with intensive ground-based programs, or from lunar orbit,
penetrating $\sim 20$~m.
Orbital missions can combine different frequencies and/or reception angles to
improve spatial resolution and ground clutter, and reduce interference speckle
noise.
Earth-based radar maps exist at 40, 430 and 800 MHz (Thompson
\& Campbell 2005), also 2.3 GHz (Stacy 1993, Campbell et al.\ 2006a, b).
Angles of incidence from Earth are large e.g., $\sim 60^\circ$, with
echoes dominated by diffuse scattering not easily modulated.
Circular polarization return can probe for surface water ice (Nozette
1996, 2001) but is questioned (Simpson 1998, Campbell et al.\ 2006a).
Applying these to subsurface ice is at least as problematic, especially
at $\sim$300 MHz to penetrate $\sim$15~m.

Finding subsurface ice is challenging.
The dielectric constant is $K \approx 3$ for regolith, water ice (slightly
higher), and many relevant powders of comparable specific gravity e.g.,
anorthosite and various basalts.
These have attenuation lengths similar to ice, as well.
Using net radar return alone, it will be difficult to distinguish ice from
regolith.
In terrestrial situations massive ice bodies reflect little internally
(Moorman, Robinson \& Burgess 2003).
Ice-bearing regions should be relatively dark in radar images, if lunar
ice-infused volumes homogenize or ``anneal,'' either forming a uniform slab or
by binding together regolith with ice in a uniform $K$ bulk.
On the other hand, hydrated regolith has $K$ much higher than unhydrated (up to
10 times), and attenuation lengths over 10 times shorter (Chung 1972).
Hydration effects are largest at lower frequencies, even below 100 MHz.
If water ice perturbs regolith chemistry, increasing charge mobility as in a
solution, $K$ and conductivity increase, raising the loss tangent (conductivity
divided by $K$ and frequency).
This high-$K$ zone should cause reflections, depending strongly on the
suddenness of the transition interface.

The 430 MHz radar map (Ghent et al.\ 2004) of Aristarchus and vicinity, site of
$\sim$50\% of TLP and radon reports shows
the 43-km diameter crater surrounded by low radar-reflectivity some
150~km across, especially downhill from the Aristarchus Plateau, which is
dark to radar, except bright craters and Vallis Schr\"oteri.
The darker radar halo centered on Aristarchus itself is uniquely smooth,
indicating that it was probably formed or modified by the impact, a few hundred
My ago.
This darkness might be interpreted as high loss tangent, as above, or simply
fewer scatterers (Ghent et al.~2004) i.e., rocks of $\sim$1m size; it is
undemonstrated why the latter applies in the ejecta blanket within the bright
radar halo within 70 km of the Aristarchus center.
Other craters, some as large as Aristarchus, have dark radar haloes, but none
so extended (Ghent et al.~2005).
The Aristarchus region matches subsurface ice redistributed by impact melt:
dark, smooth radar-return centered on the impact (although tending downslope).
One should search for dark radar areas around likely outgassing sites.

\subsubsection {Monitoring from Low Lunar Orbit}

\noindent {\it Alpha-Particle Spectrometry:}
A $^{222}$Rn atom random walks only $\sim$200~km before decaying (or sticking
to a cold surface).
In under a day, $^{222}$Rn dispersal makes superfluous placing detectors $<$100
km above the surface (excepting $r^{-2}$ sensitivity considerations).
Alpha-particle spectrometers observed the Moon successfully for short times.
The latitude coverage was limited on {\it Apollo 15} ($\mid Lat\mid\la
26^\circ$ for 145 hours) and {\it Apollo 16} ($\mid Lat\mid\la 5^\circ$, 128~h).
{\it Lunar Prospector's} Alpha Particle Spectrometer (covering the entire Moon
over 229 days spanning 16 months) was partially damaged and suffered
sensitivity drops due to solar activity (Binder 1998).
$Kaguya$'s Alpha Ray Detector (ARD) promised 25 times more sensitivity than
Apollo (Nishimura et al.\ 2006), but sharing a failed power supply it has yet
to produce results.
{\it Apollo 15} observed outgassing events from Aristarchus and Grimaldi,
{\it Apollo 16} none, and {\it Lunar Prospector} Aristarchus and Kepler
integrated over the mission.
Apollo and {\it Lunar Prospector} detected decay product $^{210}$Po
at mare/highlands boundaries from $^{222}$Rn leakage over the past $\sim$100 y.
An expected detection rate might be grossly estimated, consistent with an event
1-2 times per month detectable by {\it Apollo 15}, and by {\it Lunar
Prospector} over the mission, with Aristarchus responsible for $\sim$50\%.
A polar orbiting alpha-particle spectrometer with a lifetime of a year or more
and instantaneous sensitivity equal to Apollo's could produce a detailed map of
outgassing on the lunar surface separate from optical manifestation.
Two in polar orbit could cover the lunar surface every 1.8 half-lives of
$^{222}$Rn, nearly doubling sensitivity.
Sensitivity can be increased if detectors incorporate solar wind vetos, or
operate
during solar minimum, and if detectors orient towards the lunar surface.

\noindent
{\it On-Orbit Mass Spectrometry:}
Unlike $^{222}$Rn and its long surface residence, other outgassing events call
for several instruments for efficient localization e.g., by mass spectrometry.
With outgassing of hundreds of tons and tens of events per year, particle mass
fluence from one outburst seen 1000~km away approaches $10^{12}$~AMU~cm$^{-2}$.
A burst that is seen by a few detectors could be well constrained.
Gas scale heights $\sim$100~km imply detectors near the ground.
Conversely, an instantaneous outburst seen 100~km away will disperse
less than one minute in arrival; detectors must operate rapidly.
This was a problem e.g., the {\it Apollo 15} Orbital Mass Spectrometer
Experiment (Hoffman \& Hodges 1972) requiring 62s to scan through a factor of
2.3 in mass.
Clearly there are two separate modes of gas propagation above the
surface, neutral and ionized (Vondrak, Freeman \& Lindeman 1974, Hodges et
al.~1972), at rates of one to hundreds of tonne~y$^{-1}$ for each.

Operational strategies of these detectors are paramount.
Consider an event 1000~km away, which will spread $\sim 500$s in arrival time.
A simple gas pressure gauge is too insensitive; with an ambient
atmosphere not atypical e.g., number density $n \approx 10^4-10^5$ cm$^{-3}$
(varying day/night e.g., Hodges, Hoffman \& Johnson 2000), the collisonal
background rate in 500~s amounts to 10$\times$ or more than the fluence for a
typical outburst (assuming $\sim 20$ AMU particles).
Since interplanetary solar proton densities can vary by order unity
in an hour or less (e.g., McGuire 2006), pressure alone is insufficient.
Mass spectrometry subdivides incoming flux in mass,
but also in direction, decreasing effective background rates.

One satellite particle detector cannot distinguish episodic behavior of
outgassing versus spacecraft motion at $\sim 1.7$ km s$^{-1}$.
Localizing such signals between two platforms is ideal, at least for neutral
species, if they constrain temporal/spatial location of specific outbursts
using timing and signal strength differences.
A timing difference indicates the distance difference to the source, with the
source confined to a hyperboloid locus.
Location on this hyperboloid is fixed by signal strengths, plus
left/right ambiguity from detector directionality.
A mean nearest satellite distance of 1000~km from arbitrary
sources requires $\ga$10 low orbital platforms.
Mass spectrometers on the surface can maintain such density over smaller areas
efficiently once we know roughly where sources may be.
A mass spectrometer planned for {\it Lunar Atmospheric and Dust
Environment Explorer} ($LADEE$) sits on one platform in equatorial orbit;
geographical resolution of outgassing events will be poor.\footnote{
``Low maintenance'' lunar orbits at low altitude require few corrections due to
mascon, but must maintain
``frozen orbit'' inclination angles $i = $27, 50, 76 or $86^\circ$ (e.g.,
Ramanan \& Adimurthy 2005).
If we want to maintain a position over the terminator (sun-synchronous orbit),
we requires a precession rate $\omega_p =0.99^\circ d^{-1}=2\times 10^{-7}$ rad
s$^{-1}$.
Alternatively, LADEE achieves this by a precessing, highly eccentric orbit, but
spends a small fracton of its time near the lunar surface.
Precession is fixed by
coefficient $J_2 = (2.034 \pm 0.001) \times 10^{-4}$ (Konopliv et al.~1998)
according to $\omega_p =-(3 a^2 J_2 \sqrt{GM} cos~i)/(2 r^{7/2})$, where $a$ is
lunar radius,
$\omega$ orbital angular speed, $M$ lunar mass and $r$ orbital radius.
Precession due to the Sun and Earth are much smaller.
One cannot effectively institute both sun synchronicity in a polar orbit,
however, since the maximum
inclination orbit with $\omega_p = 2 \times 10^{-7}$ s$^{-1}$ occurs at
$47^\circ$ (or else below the surface).
To force sun-synchronicity at $i = 86^\circ$,
$\omega_p = 1.5 \times 10^{-8}$ s$^{-1}$, requires
only $a = 0.3$~mm s$^{-2}$ which could even be accomplished by a
Hall-effect ion engine or even a solar sail (with 330 cm$^2$ g$^{-1}$).}

\subsubsection{In-Situ and Near Surface Exploration}

Surveying {\it in situ} approaches to studying volatiles is beyond the scope of
this paper;\footnote{Our research group, AEOLUS: ``Atmosphere seen from Earth,
Orbit and the LUnar Surface,'' is developing ways to efficiently transfer
information from remote sensing to {\it in situ} research of lunar volatiles.}
we emphasize a few key points.
The key effort is to focus from wide-ranging reconnaissance down to scales
where lunar volatiles can be sampled near their source.
Primary global strategies are optical transient monitoring (Near Side,
resolution $\sim$1~km) and orbital $^{222}$Rn and 3$\mu$m detection (both
hemispheres, $\sim$100~km and $<$1~km, respectively).
Even trusting that TLPs trace volatiles and centroiding TLPs to 10\% of
a resolution element, localization error ($\sim$100~m) could
preclude easy {\it in situ} followup.
(In Appendix I we outline improving this to $\sim 10$~m.)
Two simple {\it in situ} technologies could isolate 
outgassing sources below 100~km scales.
First, three alpha particle detectors on the surface can triangulate nearby
$^{222}$Rn outgassing sources, using strength and time delay in arrival of
random walking $^{222}$Rn.
Secondly, a mass spectrometer that can reconstruct the ballistic trajectory of
neutrals from the source (Austin et al.\ 2008, Daly, Radebaugh \& Austin 2009)
can construct an ``image'' of transient outgassing sources over regions up
to 1000~km across.
This spectrometer is not overwhelmed by pulsed sources while measuring masses
over a wide range.
Further technologies could pinpoint subsurface
structure at 10~m scales from information at 1~km.\footnote{These
include local seismic arrays, local ground penetrating radar, magnetometer
and infrared laser arrays.
The latter would consist of lasers, mirrors and sensors on towers
for specific molecular species, either constructed to
exploit a specific species e.g., CO$_2$, or tuned to one of several
species' vibrational-rotational states.
On smaller scales (1-100~m) several varieties of mass spectroscopy might
prove effective, including downward-sniffing spectrometers, triangulating
outburst detectors arrays, and pyrolysis mass spectrometers (ten Kate et
al.\ 2009) which heat regolith samples in search of absorbed species from
previous outgassing.}

By LACE's deployment with the final Apollo landing, the outgassing environment
was contaminated by anthropogenic gas (Freeman \& Hills 1991) 
especially near landing sites; each mission of human exploration will deliver
tens of tonnes gases to the surface, with species relevant to endogenous
volcanic gas, approaching or exceeding the annual endogenous output of such
gases.\footnote{
Constellation spacecraft Orion burn N$_2$O$_4$ (nitrogen teroxide) and
CH$_3$N$_2$H$_3$ (mono-methyl hydrazine), with Altair propelled by
liquid oxygen and hydrogen.
Future missions might use liquid O$_2$ and CH$_4$.
Earth Departure Stages might deliver residual $O_2$ and $H_2$ in lunar impact.
Altair (and EDS) produce water, and Orion exhausts H$_2$O, CO$_2$ and N$_2$.
N$_2$ was the prime candidate constituent in an outburst seen by the
{\it Apollo 15} over Mare Orientale: Hoffman \& Hodges 1972, perhaps
anthropogenically - Hodges 1991.}
Depending on spacecraft orientations and trajectories when thrusting,
they may deliver $\sim$20 tonnes of mostly water to the surface, which
will remain up to about one lunation, making suspect measurements of these and
other species for years.


\section{Discussion and Conclusions}

The origin of TLPs has been mysterious, and their
correlation to outgassing, while strong, was only circumstantial.
The plausible generation of TLP-like events as simple consequence
of outgassing from the interior lends credence to a possible causal link.
We present a model tied to outgassing from deep below
the regolith that reproduces the time and spatial scale of reported TLPs,
suggesting a causal link to outgassing.
Radiogenic gas evolved from the regolith cannot
provide the concentration to produce a noticeable explosive event.

Apollo and later data were insufficiently sensitive to establish the level
of outgassing beyond $^{222}$Rn, and isotopes of Ar, plus He,
presumably, and did detect marginally molecular gas, but of uncertain origin,
particularly CH$_4$,
Reviewing the evidence and available techniques, there are several gases that
should be highlighted as crucial outgassing tracers.
$^{222}$Rn (and its products e.g., $^{210}$Po) can be detected remotely of
course and are unique in terms of their mapping potential, while being a minor
fraction of escaping gas, presumably.
$^{40}$Ar is a major mass constituent of the atmosphere and unlike $^4$He is
not confused with the solar wind.
Both $^{222}$Rn and $^{40}$Ar will favor KREEP terrane in the western
maria, presumably.

If outgassing arises in the deep interior, one cannot neglect
indications that at one time this was dominated by volcanic, molecular
gas.
Particular among these is water vapor, passing its triple point
temperature in rising through the regolith.
Given a high enough concentration, therefore, one should expect the production
of water ice.
The conditions under the regolith, particularly near the lunar poles, are
favorable for such ice to persist even over geological time interval.
It is possible that ice generated there when outgassing was more active still
remains.
We further point out that the plausible chemical interaction of such molecular
gases with the regolith is the production of cement-like compounds that might
radically alter the diffusivity of the regolith.
Given the temperature evolution of the regolith, this non-diffusive layer would
isolate the volatile outflow from the vacuum.

The question remains how we will detect such molecular outgassing effects,
given their largely covert nature; this is greatly complicated by possible
anthropogenic contamination in the future.
Among molecular gases, sulfur e.g., SO$_2$ is the predominate volatile detected
in deep-interior fire fountain glasses, a key factor in deciding what to pursue
as a volcanic tracer.
Furthermore, essential no liquid or hybrid rocket propellant candidate contain
sulfur,
and the only sulfuric solid propellants are fairly outdated e.g., black powder
and Zn-S.
NASA and hopefully other space agencies have no plans to use these on lunar
missions.

Despite the hypotheses and methods outlined above, there is great doubt
regarding the nature of lunar outgassing.
Water is of obvious and diverse interest, and CO$_2$ and CO, while missing as
apparent constituents, are interesting as drivers for fire fountain eruption.
Plausibly the only way to study these components reliably is before the new
introduction of large spacecraft into the lunar environment.
Given uncertainty of how these gases and SO$_2$ might interact with the
regolith, this early study appears paramount.

Significantly, many years to come
monitoring for optical transients will be best done from Earth's
surface, even considering the important contributions that will be made by
lunar spacecraft probes in the near future.
However, these spacecraft will be very useful in evaluating the nature of
transient events in synergy with ground-based monitoring.
Given the likely behavior of outgassing events, it is unclear that in-situ
efforts alone will necessarily isolate their sources within significant
winnowing of the field by remote sensing.
Early placement of capable mass spectrometers of the lunar surface, however,
might prove very useful in refining our knowledge of outgassing composition,
in particular a dominant component that could be used as a tracer to monitor
outgassing activity with more simple detectors.
This should take place before significant atmospheric pollution by large
spacecraft, which will produce many candidate tracer gases in their exhaust.

Finally, as we edit this paper's final version, several works have become
available indicating confirmed lunar regolith hydration signals
(Pieters et al.\ 2009, Clark 2009, Sunshine et al.\ 2009) in the 3$\mu$m band,
and we comments about these here.
These show a strong increase in hydration signal towards the poles, as
predicted in \S 4.
To our knowledge our model is uniquely consistent with this and the
general hydration signal strength, in places $\ga$700 ppm by mass
(also with Vilas et al.\ 1999, 2008).
Unfortunately the Moon Mineralogy Mapper (M$^3$: Pieters et al.\ 2006), in
finding this signal, but not completely mapping it, provided
tentative indication of its large scale distribution varying over a lunation.
This variation can be studied from Earth with a simple near/mid IR
camera (InSb or red-extended HgCdTe) with on- and off-band filters for IR
hydration bands (or 0.7$\mu$m: Vilas et al.\ 1999).
To complete this valuable work, along with other instruments
needing a lunar polar orbiter (alpha-particle spectrometer,
ground-penetrating radar, mass spectrometers, etc.), an instrument similar to
M$^3$ should probably fly again before human lunar missions.
Note that the same type and level of signal was detected by {\it Luna 24}
(Akhmanova et al.\ 1978), and these authors believed it not due to
terrestrial contamination.
They detected increasing hydration with depth into the regolith,
a likely circumstance in our model.
This core sample reached 2 meters depth, several times deeper than
epithermal neutrons e.g., seen on {\it Lunar Prospector} or {\it
LRO}, and corresponding to an impact gardening over $\sim$2~G.y.
Such a gradient arises naturally from seepage of
water vapor, but water and/or hydroxyl from solar wind
proton implantation may not explain the
concentration of 3$\mu$m signal to the poles and
the {\it Luna 24} hydration depth profile.
This offers a challenge for this model, or for water delivered by comets and/or
meteoroids.
In a separate paper we will review further evidence supporting endogenous
origin of lunar hydration.

\newpage
\begin {center}
{\bf Acknowledgements}
\end {center}

We would much like to thank Alan Binder and James Applegate, as well as Daniel
Savin, Daniel Austin, Ed Spiegel and the other members of AEOLUS (``Atmosphere
as seen from, Earth, Orbit and LUnar Orbit'') for helpful discussion.
This research was supported in part by NASA (07-PAST07-0028 and
07-LASER07-0005), the National Geographic Society (CRE Grant 8304-07), and
Columbia University.

\bigskip

\bigskip

\bigskip

\begin{center}
{\bf Appendix I: Imaging from High Orbit}:
\end{center}

Given constraints on imaging from Earth, we
consider imaging monitors closer to the Moon.
We propose no special-purpose missions, but
detectors that could ride on other platforms e.g., 
does lunar exploration require
communications with line-of-sight access to all
points on the Moon's surface (except within deep craters, etc.)?
This might also serve for comprehensive imaging monitoring.
A minimal full network has a tetrahedral geometry with points
$\sim$60000~km above the surface: a single platform
at Earth-Moon Lagrange point L1, covering most of the Near Side, and three
points in wide halo orbits around L2 for
the Far Side plus limb seen from Earth.
Proposals exist for an L1 orbital transfer facility (Lo 2004, Ross 2006).
No single satellite sees the entire Far Side, especially since
farside radio astronomy might restrict low-frequency transmission i.e.,
lasers only.
One L2 satellite covers $\la$97\% of the Far Side (subtending
$176^\circ .8$, selenocentrically); full coverage (plus some
redundancy) requires three satellites (plus L1).
With this configuration, the farthest point from a satellite will be
typically $71^\circ$ (selenocentrically), foreshortened by $\sim 3$ times.
Such an imaging monitor might be ambitious;
to achieve 100m FWHM at the lunar sub-satellite point
requires $\sim$4~Gpixels, aperture $\ga 0.5$~m,
and field-of-view 3$^\circ .3$.
Each such monitor on an existing platform will cost perhaps \$100M.
In the meantime, we should accomplish what we can from the ground. 

\newpage

\noindent
{\bf References:}

\noindent
Adams, J.B.\ 1974, JGR, 79, 4829.

\noindent
Adorjan, A.S.\ 1970, J.\ Spacecraft \& Rockets, 7, 378.

\noindent
Akhmanova, M.V., Dement'yev, B.V.\ \& Markov, M.N.\ 1978, Geokhimiya, 2, 285.

\noindent
Akhmanova, M.V., Dement'yev, B.V., Markov, M.N.\ \& Sushchinskii, M.M.\ 1972,
Cosmic Res., 10, 381

\noindent
Allen, C.C., et al.\ 1998, Lun.\ Plan.\ Sci.\ Conf., 29, 1690.

\noindent
Anders, E.\ 1970, Science, 169, 1309.

\noindent
Anders, E. Ganapathy, R., Kr\"ahenb\"uhl, U.\ \& Morgan, J.W.\ 1973, The Moon,
8, 3.

\noindent
Anderson, R.. et al.\ 1965, Science, 148, 1179.

\noindent
Andreas 2007, Icarus, 286, 24.

\noindent
Austin, D.E., Miller, I., Daly, T., Crotts, A., Syrstad, E., Brinckerhoff,
W.\ \& Radebaugh, J.\ 2008, NLSI Lun.\ Sci.\ Conf., 2074.

\noindent
Bart, G.D.\ \& Melosh, H.J.\ 2005, D.P.S., 57, 57.07.

\noindent
Bautsch, M., Varadinek, P., Wege, S.\ \& Niedrig, H.\ 1994, 
J.\ Vac.\ Sci.\ Technol.\ A., 12, 591.

\noindent
Basu, A.\ \& Molinaroli, E.\ 2001, Earth Moon \& Plan., 85, 25.

\noindent
Brantley, S.L.~2004, in ``Treatise on Geochemistry'' eds.~H.D.~Holland \&
K.K.~Turekian (Elsevier: Amsterdam), section 5.03.

\noindent
Brown, W.E., Jr.\ 1972, Earth Moon Plan., 4, 133.

\noindent
Buratti, B.J., McConnochie, T.H., Calkins, S.B., Hillier, J.K.\ \&
Herkenhoff, K.E.\ 2000, Icarus, 146, 98.

\noindent
Bussey, B., Spudis, P.D., Lichtenberg, C., Marinelli, B.\ \& Nozette,
S.\ 2006, in {\it LCROSS Selection Conf.}, (LPI: Houston), 9013.

\noindent
Cameron, W.S.\ 1972, Icarus, 16, 339.

\noindent
Cameron, W.S.\ 1977, Phys.\ Earth Planet.\ Inter., 14, 194.

\noindent
Campbell, B.A., Campbell, D.B., Margot, J.-L., Ghent, R.R., Nolan, M., Carter,
L.M., Stacy, N.J.S.\ 2007, Eos, 88, 13.

\noindent
Campbell, D.B., Campbell, B.A., Carter, L.M., Margot, J.-L.\ \& Stacy,
N.J.S.\ 2006a, Nature, 443, 835.

\noindent
Campbell, B.A., Carter, L.M., Campbell, D.B., Hawke, B.R., Ghent, R.R.\ \&
Margot, J.-L.\ 2006b, Lun.\ Plan.\ Sci.\ Conf., 37, 1717.

\noindent
Carbognani, A.\ 2004, Astronomia, 5, 12.

\noindent
Carrier, W.D., Olhoeft, G.R.\ \& Mendell, W.\ 1991, in ``Lunar Sourcebook,''
eds.~G.H.\ Heiken, D.T.\ Vaniman \& B.M.\ French (Cambridge U.: Cambridge),
p.\ 475.

\noindent
Charette, M.P., Adams, J.B., Soderblom, L.A., Gaffey, M.J.\ \& McCord,
T.B.\ 1976, Lun.\ Sci.\ Conf., 7, 2579.

\noindent
Charles, R.W., Hewitt, D.A.\ \& Wones, D.R.\ 1971, Lun.\ Sci.\ Conf., 1, 645.

\noindent
Chevrier, V., et al.\ 2007, GRL, 34, L02203.

\noindent
Chin, G., et al.\ 2007, Lun.\ Plan.\ Sci.\ Conf., 38, 1764.

\noindent
Chung, D.H.\ 1972, Earth Moon \& Plan., 4, 356.

\noindent
Clark, R.N.\ 2009, Science Express Rep., 10.1126/science.1178105.

\noindent
Cocks, F.H., et al.\ 2002, Icarus, 160, 386.

\noindent
Collins, G.S.\ 2001, Lun.\ Plan.\ Sci.\ Conf., 32, 1752.

\noindent
Compton, K.T., Lilly, E.G.\ \& Olmstead, P.S.\ 1920, Phys.\ Rev., 16, 282.

\noindent
Crotts et al. 2009, in preparation.

\noindent
Crotts, A.P.S.\ 2008, ApJ, 687, 1186.

\noindent
Crotts, A.P.S.\ 2009, ApJ, 697, 1.

\noindent
Cullingford, H.S.\ \& Keller, M.D.\ 1988, in {\it 2nd Conf.\ on Lunar Bases and
Space Activities}, ed.\ W.W.\ Mendell, NASA Conf.\ Pub.\ 3166, p.\ 497.

\noindent
Daly, T., Radebaugh, J.\ \& Austin, D.E.\ 2009, Lun.\ Plan.\ Sci.\ Conf., 40,
2411.

\noindent
Davis, S.S.\ 2009, Icarus, 202, 383.

\noindent
Dollfus, A.\ 2000, Icarus, 146, 430.

\noindent
Drake, M.J.\ \& Stimpfl, M.\ 2007, Lun.\ Plan.\ Sci.\ Conf., 38, 1179.

\noindent
Dzhapiashvili, V.P.\ \& Ksanfomaliti, L.V.\ 1962, The Moon, IAU Symp.\ 14,
(Academic Press: London), p.\ 463.

\noindent
Elkins-Tanton, L.T., Chatterjee, N.\ \& Grove, T.L.\ 2003, Meteor.\ \&
Plan.\ Sci., 38, 515.

\noindent
Epstein, S.\ \& Taylor, H.P., Jr.\ 1972, Geochim.\ Cosmochim.\ Acta, 36
(Suppl.\ 3), 1429.

\noindent
Farmer, C.B.\ 1976, Icarus, 28, 279.

\noindent
Farr, T.G., Bates, B., Ralph, R.L.\ \& Adams, J.B.\ 1980, Lun.\ Plan.\ Sci.,
11, 276.

\noindent
Feynmann, R.P.\ 1974, Engineer.\ \& Sci., 37, 7.

\noindent
Freeman, J.W., Jr.\ \& Hills, H.K.\ 1991, GRL, 18, 2109.

\noindent
Fried, D.L.\ 1978, Opt.\ Soc.\ Am.\ J., 68, 1651.

\noindent
Friedman, B., Saal, A.E., Hauri, E.H., van Orman, J.A.\ \& Rutherford,
M.J.\ 2009, Lun.\ Plan.\ Sci.\ Conf., 40, 2444.

\noindent
Friesen, L.J.\ 1975, The Moon, 13, 425.

\noindent
Gammage, R.B.\ \& Holmes, H.F.\ 1975, Lun.\ Sci.\ Conf., 6, 3305.

\noindent
Garlick, G.F.J., Steigmann, G.A.\ \& Lamb, W.E.\ 1972a, Nature, 238, 13.

\noindent
Garlick, G.F.J., Steigmann, G.A., Lamb, W.E.\ \& Geake, J.E.\ 1972b, 
1972, Lun.\ Plan.\ Sci.\ Conf., 3, 2681.

\noindent
Garvin, J., Robinson, M., Skillman, D., Pieters, C., Hapke, B.\ \& Ulmer,
M.\ 2005, $HST$ Proposal GO 10719.

\noindent
Gault, D.E., H\"orz, F, Brownlee, D.E.\ \& Hartung, J.B.\ 1974,
Lun.\ Plan.\ Sci.\ Conf., 5, 260.

\noindent
Geake, J.E.\ \& Mills, A.A.\ 1977, Phys.\ Earth Planet.\ Inter., 14, 299.

\noindent
Gerlach, T.M.\ \& Graeber, E.J.\ 1985, Nature, 313, 274

\noindent
Ghent, R.R., Leverington, D.K., Campbell, B.A., Hawke, B.R.\ \& Campbell,
D.B.\ 2004, Lun.\ Plan.\ Sci.\ Conf., 35, 1679.

\noindent
Ghent, R.R., Leverington, D.K., Campbell, B.A., Hawke, B.R.\ \& Campbell,
D.B.\ 2005, JGR. 110, doi: 10.1029/2004JE002366.

\noindent
Gibson, E.K.\ \& Moore, G.W.\ 1973, Science, 179, 69.

\noindent
Haynes, D.R., Tro, N.J.\ \& George, S.M.\ 1992, J.\ Phys.\ Chem., 96, 8502.

\noindent
Hazen, R.M., Bell, P.M.\ \& Mao, H.K.\ 1978, Lun.\ Plan.\ Sci., 9, 483.

\noindent
Hodges, R.R., Jr.\ 1977, Phys.\ Earth \& Planet.\ Inter., 14, 282.

\noindent
Hodges, R.R., Jr.\ 1980, Proc.\ Lun.\ Plan.\ Sci.\ Conf., 11, 2463.

\noindent
Hodges, R.R., Jr.\ 1991, {\it personnel communication}, in Stern, A.\ 1999,
Rev.\ Geophys., 37, 4.

\noindent
Hodges, R.R., Jr., Hoffman, J.H., Yeh, T.T.J.\ \& Chang, G.K.\ 1972, JGR, 77,
4079

\noindent
Hodges, R.R., Jr., Hoffman, J.H., Johnson, F.S.\ \& Evans, D.E.\ 1973,
Lun.\ Sci.\ Conf., 4, 2855.

\noindent
Hodges, R.R., Jr., Hoffman, J.H.\ \& Johnson, F.S.\ 1974, Icarus, 21, 415.

\noindent
Hodges, R.R., Jr.\ \& Hoffman, J.H.\ 1975, Lun.\ Plan.\ Sci.\ Conf., 6, 3039.

\noindent
Hoffman, J.H.\ \& Hodges, R.R., Jr.\ 1975, Moon, 14, 159.

\noindent
Horiguchi, T., Saeki, N., Yoneda, T., Hoshi, T.\ \& Lin, T.D.\ 1996, in ``Space
V: 5th Internat'l Conf.\ Engin., Constr.\ \& Operat.\ in Space,''
ed.\ S.W.\ Stewart, ASCE Proc., 207, 86.

\noindent
Horiguchi, T., Saeki, N., Yoneda, T., Hoshi, T.\ \& Lin, T.D.\ 1998, in ``Space
`98: 6th Internat'l Conf.\ Engin., Constr.\ \& Operat.\ in Space,''
eds.\ R.G.\ Galloway \& S.L.\ Lokaj, ASCE Proc., 206, 65.

\noindent
Hughes, D.W.\ 1980, Nature, 285, 438.

\noindent
Ingersoll, A.P.\ 1970, Science, 168, 972.

\noindent
Kanamori, H.\ in {\it Concrete Under Severe Conditions 1: Environment and
Loading}, eds.\ K.\ Sakai, N.\ Banthia \& O.E.\ Gjorv (Chapman \& Hall:
London), p.\ 1283.

\noindent
Kozyrev, N.A.\ 1958, Sov.\ Intern't'l Geop.\ Yr.\ Bull., PB 13162-42
(see also 1962, in {\it The Moon, IAU Symp.\ 14}, eds.\ Z.\ Kopal \&
Z.K.\ Mikhailov (Academic: New York), p.\ 263.

\noindent
Kozyrev, N.A.\ 1963, Nature, 198, 979.

\noindent
Kr\"ahenb\"uhl, U., Ganapathy, R., Morgan, J.W.\ \& Anders, E.\ 1973, Science,
180, 858.

\noindent
Langseth, M.G., Jr., Clarke, S.P., Jr., Chute, J.L., Jr., Keihm, S.J.\ \&
Wechsler, A.E.\ 1972, in {\it Apollo 15 Preliminary Science Report}, NASA
SP-289, p.\ 11-1.

\noindent
Langseth, M.G., Jr., Keihm, S.J.\ \& Chute, J.L., Jr.\ 1973, in {\it Apollo
17 Preliminary Science Report}, NASA SP-330, p.\ 9-1.

\noindent
Langseth, M.G.\ \& Keihm, S.J.\ 1977, in {\it Soviet-American Conference on
Geochemistry of the Moon and Planets}, NASA SP-370, p.\ 283.

\noindent
Law, N.M., Mackay, C.D.\ \& Baldwin, J.E.\ 2006, A\&A, 446, 739.

\noindent
Lebofsky, L.A., Feierberg, M.A., Tokunaga, A.T., Larson, H.P.\ \& Johnson,
J.R.\ 1981, Icarus, 48, 453

\noindent
Lipsky, Yu.N.\ \& Pospergelis, M.M.\ 1966, Astronomicheskii Tsirkular, 389, 1.

\noindent
Lo, M.W.\ 2004, in ``Proc.\ Internat'l Lunar Conf.\ 2003, ILEWG 5'' (Adv.\ in
Astronaut.\ Sci., Sci.\ \& Tech. Ser., Vol.\ 108), eds.\ S.M.\ Durst et al.
(Univelt: San Diego), p.\ 214.

\noindent
Lucey, P.G., Blewett, D.T., Taylor, G.J.\ \& Hawke, B.R.\ 2000, JGR, 105, 20377.

\noindent
Lucey, P.G., Taylor, G.J.\ \& Hawke, B.R.\ 1998, Lun.\ Plan.\ Sci.\ Conf., 29,
1356.

\noindent
Lunine, J., Graps, A., O'Brien, D.P., Morbidelli, A., Leshin, L.\ \& Coradini,
A.\ 2007, Lun.\ Plan.\ Sci.\ Conf., 28, 1616.

\noindent
Markov, M.N., Petrov, V.S., Akhmanova, M.V.\ \& Dementev, B.V.\ 1979, in {\it
Space Research, Proc.\ Open Mtgs.\ Working Groups} (Pergamon: Oxford), p.\ 189.

\noindent
Martin, R.T., Winkler, J.L., Johnson S.W.\ \& Carrier, III, W.D.\ 1973,
``Measurement of conductance of Apollo 12 lunar simulant taken in the molecular
flow range for helium, argon, and krypton gases.'' Unpublished report quoted in
Carrier et al.\ (1991).

\noindent
McCubbin, F.M., Nekvasil, H.\ \& Lindsley, D.H.\ 2007, Lun.\ Plan.\ Sci.\ Conf.,
38, 1354.

\noindent
McEwen, A.S., Robinson, M.S., Eliason, E.M., Lucey, P.G., Duxbury, T.C.\ \&
Spudis, P.D.\ 1994, Science, 266, 1858.

\noindent
McKay, D.S., Fruland, R.M.\ \& Heiken, G.H.\ 1974, Lun.\ Plan.\ Sci.\ Conf., 5,
887.

\noindent
McKay, D.S., et al.\ 1991, in ``Lunar Sourcebook,'' eds.\ G.H.\ Heiken,
D.T.\ Vaniman \& B.M.\ French (Cambridge U.\ Press: Cambridge), p.~285.

\noindent
Middlehurst, B.M.\ 1977, Roy.\ Soc.\ Phil.\ Trans.\ A, 285, 485.

\noindent
Mills, A.A.\ 1969, Nature, 224, 863

\noindent
Mills, A.A.\ 1970, Nature, 225, 939.

\noindent
Moorman, B.J., Robinson, S.D.\ \& Burgess, M.M.\ 2003, Permafrost \&
Periglac.\ Proc., 14, 319.

\noindent
Morgan, T.H.\ \& Shemansky, D.E.\ 1991, JGR, 96, 1351.

\noindent
Mukherjee, N.R.\ 1975, The Moon, 14, 169.

\noindent
Mukherjee, N.R.\ \& Siscoe, G.L.\ 1973, JGR, 78, 1741.

\noindent
Neukum, G., et al.\ 2001, Space \& Sci.\ Rev., 96, 55.

\noindent
Nishimura, J., et al.\ 2006, Adv.\ in Space Res., 37, 3.

\noindent
Nozette, S.\ et al.\ 1996, Science, 274, 1495.

\noindent
Nozette, S.\ et al.\ 2001, JGR, 106, 23253.

\noindent
O'Hara, M.J.\ 2000, J.\ Petrology, 41, 11, 1545.

\noindent
Ohtake, M.\ et al.\ 2007, Lun.\ Plan.\ Sci.\ Conf., 38, 1829.

\noindent
Ono, T.\ \& Oya, H.\ 2000, Earth Plan.\ Space, 52, 629.

\noindent
Pearse, R.W.B.\ \& Gaydon, A.G.\ 1963, {\it The Identification of Molecular
Spectra}, (Ugsman \& Hall: London)

\noindent
Perel, J., Mahoney, J.F., Kalensher, B.E.\ \& Forrester, A.T.\ 1981, 
Appl.\ Phys.\ Let., 38, 320.

\noindent
Pieters, C.M., et al.\ 2005, 
http://moonmineralogymapper.jpl.nasa.gov/SCIENCE/Volatiles/

\noindent
Pieters, C.\ et al.\ 2006, Lun.\ Plan.\ Sci.\ Conf., 37, 1630.

\noindent
Pieters, C.\ et al.\ 2009, Science Express Rep., 10.1126/science.1178658.

\noindent
Porcello, L.J, et al.\ 1974, Proc.\ IEEE, 62, 769.

\noindent
Quaide, W.\ \& Oberbeck, V.\ 1975, Moon, 13, 27.

\noindent
Reader, J.\ \& Corliss, C.H.\ 1980, CRC Handbook of Chemistry \& Physics, 68.

\noindent
Rivkin, A.S., Howell, E.S., Britt, D.T., Lebofsky, L.A., Nolan, M.C.\ Branston,
D.D.\ 1995, Icarus, 117, 90

\noindent
Rivkin et al. 2002, Asteroids III, 237

\noindent
Robinson, J.H.\ 1986, JBAA, 97, 12.

\noindent
Robinson, M.S.\ et al.\ 2005, Lun.\ Plan.\ Sci.\ Conf., 36, 1576.

\noindent
Ross, S.D.\ 2006, Am.\ Sci., 94, 230.

\noindent
Rubey, W.W.\ 1964, in ``Origin \& Evolution of Atmospheres \& Oceans,''
eds.\ P.J.~Brancazio \& A.G.W.~Cameron (Wiley: New York), p.~1.

\noindent
Rutherford, M.J.\ \& Papale, P.\ 2009, Geology, 37, 219.

\noindent
Saito, Y., Tanaka, S., Takita, J., Horai, K.\ \& Hagermann, A.\ 2007, Lun.\ 
Plan.\ Sci.\ Conf., 38, 2197.

\noindent
Saal, A.E., Hauri, E.H., Cascio, M.L., van Orman, J.A., Rutherford, M.J.\ \&
Cooper, R.F.\ 2008, Nature, 454, 192.

\noindent
Sato, M.\ 1976, Proc.\ Lun.\ \& Plan.\ Sci.\ Conf., 7, 1323.

\noindent
Sato, M.\ 1979, Proc.\ Lun.\ \& Plan.\ Sci.\ Conf., 10, 311.

\noindent
Schorghofer, N.\ \& Taylor, G.J.\ 2007, JGR, 112, E02010.

\noindent
Schultz, P.H., Staid, M.I.\ \& Pieters, C.M.\ 2006, Nature, 444, 184.

\noindent
Schumm, S.A.\ 1970, Geol.\ Soc.\ Amer.\ Bull., 81, 2539.

\noindent
Shearer, C.K., Layne, G.D. \& Papike, J.J.\ 1994, Geochim.\ CosmoChim.\ Acta,
58, 5349.

\noindent
Siegal, B.S.\ \& Gold, D.P.\ 1973, Moon, 6, 304.

\noindent
Simpson, R.A.\ 1998, in ``Workshop on New Views of the Moon,''
eds.\ B.L.\ Jolliff \& G.\ Ryder (LPI: Houston), p.\ 61.

\noindent
Speedy, R.J., Denebetti, P.G., Smith, R.S., Huang, C.\ \& Kay, B.D.\ 1996,
J.~Chem.~Phys., 105, 240.

\noindent
Spohn, T, Konrad, W., Breuer, D.\ \& Ziethe, R.\ 2001, Icarus, 149, 54.

\noindent
Spudis, P.H.\ \& Schultz, P.D.\ 1983, Nature, 302, 233.

\noindent
Stacy, N.J.S.\ 1993, Ph.D.\ thesis (Cornell U.).

\noindent
Stern, A.\ 1999, Rev.\ Geophys., 37, 4.

\noindent
Stimpfl, M., de Leeuw, N.H., Drake, M.J.\ \& Deymier, P.\ 2007, 
Lun.\ Plan.\ Sci.\ Conf., 38, 1183.

\noindent
Stubbs, V.J., Vondrak, R.R.\ \& Farrell, W.H.\ 2005, Lun.\ Plan.\ Sci.\ Conf.,
36, 1899.

\noindent
Sunshine, J.M.\ et al.\ 2009, Science Express Rep., 10.1126/science.1179788.

\noindent
Taylor, S.R.\ 1975, {\it Lunar Science: a Post-Apollo View} (Pergamon: NY),
372 pp.

\noindent
ten Kate, I.L., Glavin, D.P.\ \& the VAPoR team, Lun.\ Plan.\ Sci.\ Conf., 40,
2232.

\noindent
Thomas, G.E.\ 1974, Science, 183, 1197.

\noindent
Thomas, R.J., et al.\ 2007, Science, 315, 1097.

\noindent
Thompson, T.W.\ \& Campbell, B.A.\ 2005, Lun.\ Plan.\ Sci.\ Conf., 36, 1535.

\noindent
Tomaney, A.\ \& Crotts, A.P.S.\ 1996, AJ, 112, 2872.

\noindent
Tubbs, R.N.\ 2003, Ph.D.\ thesis (University of Cambridge).

\noindent
Vilas, F., Dominque, D.L., Jensen, E.A., McFadden, L.A., Coombs, C.R.\ \&
Mendell, W.W.\ 1999, Lun.\ Plan.\ Sci.\ Conf., 30, 1343.

\noindent
Vilas, F., Jensen, E.A., Dominque, D.L., McFadden, L.A., Runyon, C.J.\ \&
Mendell, W.W.\ 2008, Earth, Planets \& Space, 60, 67.

\noindent
Volquardsen, E.L., Rivkin, A.S.\ \& Bus, S.J.\ 2004, DPS, 36, 32.13.

\noindent
Vondrak, R.R., Freeman, J.W.\ \& Lindeman, R.A.\ 1974,
Lun.\ \& Plan.\ Sci.\ Conf., 5, 2945.

\noindent
Washburn, E., et al., eds.\ 2003, {\it International Critical Tables of
Numerical Data, Physics, Chemistry and Technology} (Knovel: Norwich, NY),
elec.\ edition.

\noindent
Williams, R.J.\ \& Gibson, E.K.\ 1972, Earth Plan.\ Sci.\ Let., 17, 84.

\noindent
Wilson, L.\ \& Head, J.W.\ 2003, Geophys.\ Res.\ Let., 30, 1605.

\noindent
Yue, Z., Xie, H.. Liu, J.\ \& Ouyang, Z.\ 2007, Lun.\ Plan.\ Sci.\ Conf., 38,
2082.

\noindent
Zito, R.R.\ 1989, Icarus, 82, 419.

\begin{figure}
\plotfiddle{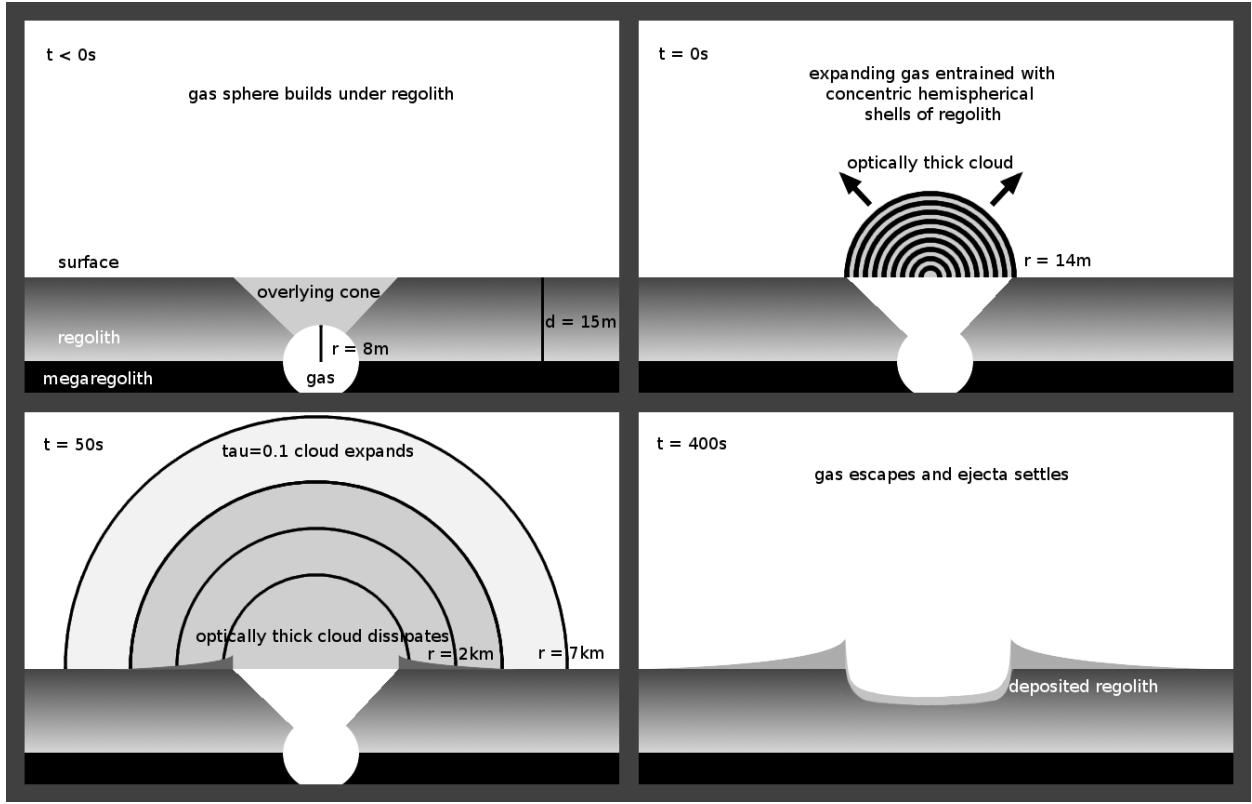}{5.0in}{000}{035}{035}{-237}{+020}
\vskip 0.00in
\caption {
A sketch of our 1D spherically-symmetric model explosion.
{\bf Frame 1:} venting gas builds up at the interface between the megaregolith and the
low-diffusivity regolith.  It builds until its pressure is sufficient to lift
the overlying cone of regolith to the surface. 
{\bf Frame 2:} we assume that this
regolith/gas mixture now translates into a hemispherical volume on the surface. 
This volume is uniform in gas density, but it is populated with concentric,
hemispherical shells of regolith particles (represented as points 45 degrees up the
side of each arc), each feeling a pressure from the expanding gas. 
As the cloud expands, the equations of motion are numerically integrated for
each shell in the cloud according to its mass and regolith particle size composition. 
{\bf Frame 3:} the
cloud expands until it reaches a point where it is no longer optically thick as
viewed from above, but the dust still entrained in the cloud continues to expand and
a thinner component ($\tau = 0.1$) remains somewhat visible from above. 
Figure 2 shows
the detailed evolution of these clouds.  Regolith that has fallen out of the gas
cloud begins to pile up on the ground around the crater. 
{\bf Frame 4:} all of the gas
has escaped and all entrained regolith is deposited on the ground around the crater
in a manner shown in Figure 4.  
}
\end{figure}

\begin{figure}
\plotfiddle{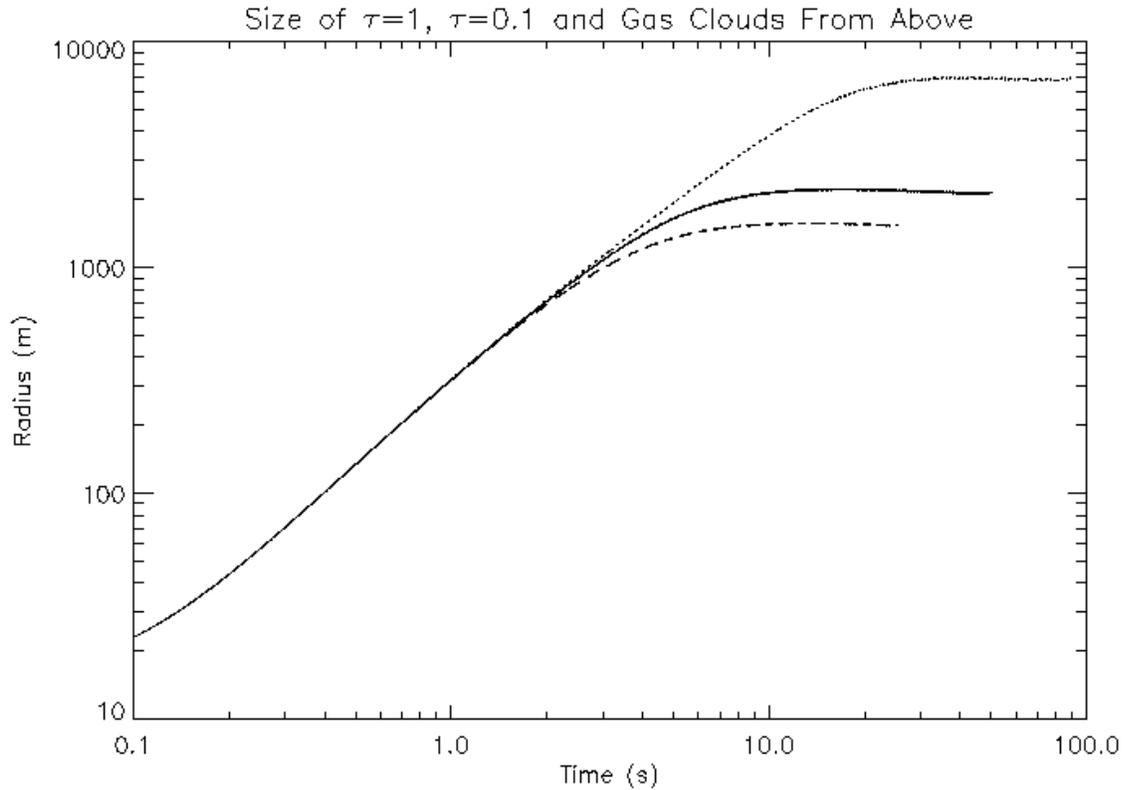}{5.0in}{000}{085}{085}{-257}{-020}
\vskip 0.00in
\caption {
The size of the various components of the clouds as seen from above, versus
time after the initial explosion, for a ``minimal TLP.''
The solid line represents the radius of the dust cloud with optical thickness
$\tau = 1$ as seen from above.
The dotted line represents the radius of the dust cloud with $\tau = 0.1$
as seen from above.
These two optical depth values correspond roughly to the level of change in
contrast that might be considered easy versus marginal to detect with the human
eye observing lunar surface features affected by the event.
The dashed line represents the radius of the outermost layer of the gas cloud
still constrained by the dust cloud.
At 10 s after the explosion, the $\tau = 1$ portion of the cloud has expanded
to 2 km radius and stops expanding; after 50 s no portion of the cloud remains
with $\tau=1$.
After 90 s, the $\tau = 0.1$ portion of the cloud has also disappeared, having
first expanded to 7 km in radius.
}
\end{figure}
 
\begin{figure}
\plotone{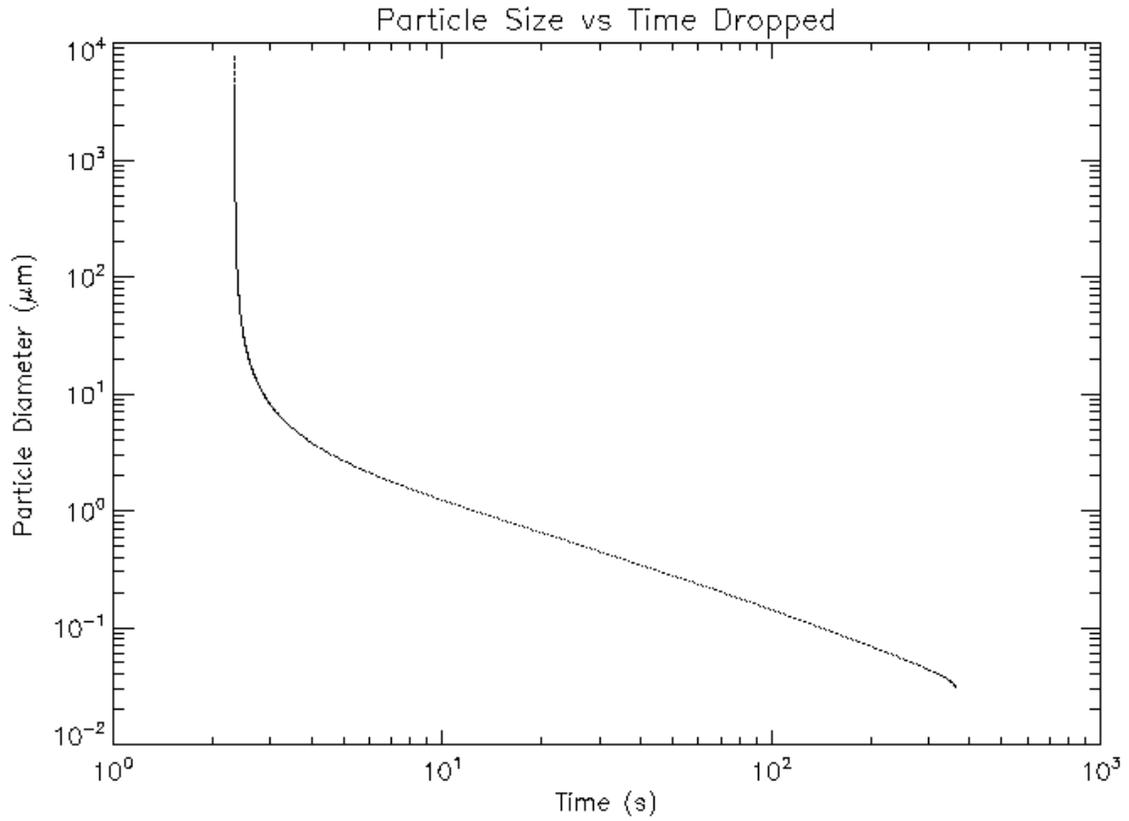}
\caption{
The time required for a typical regolith particle of the diameter shown to
fall out of the expanding cloud back to the lunar surface.
Particles larger than about 30 micron (slightly more than 50\% by mass) rain
immediately to the ground, whereas particles smaller than optical wavelength
remain aloft for at least several minutes.
}
\end{figure}

\begin{figure}
\plotone{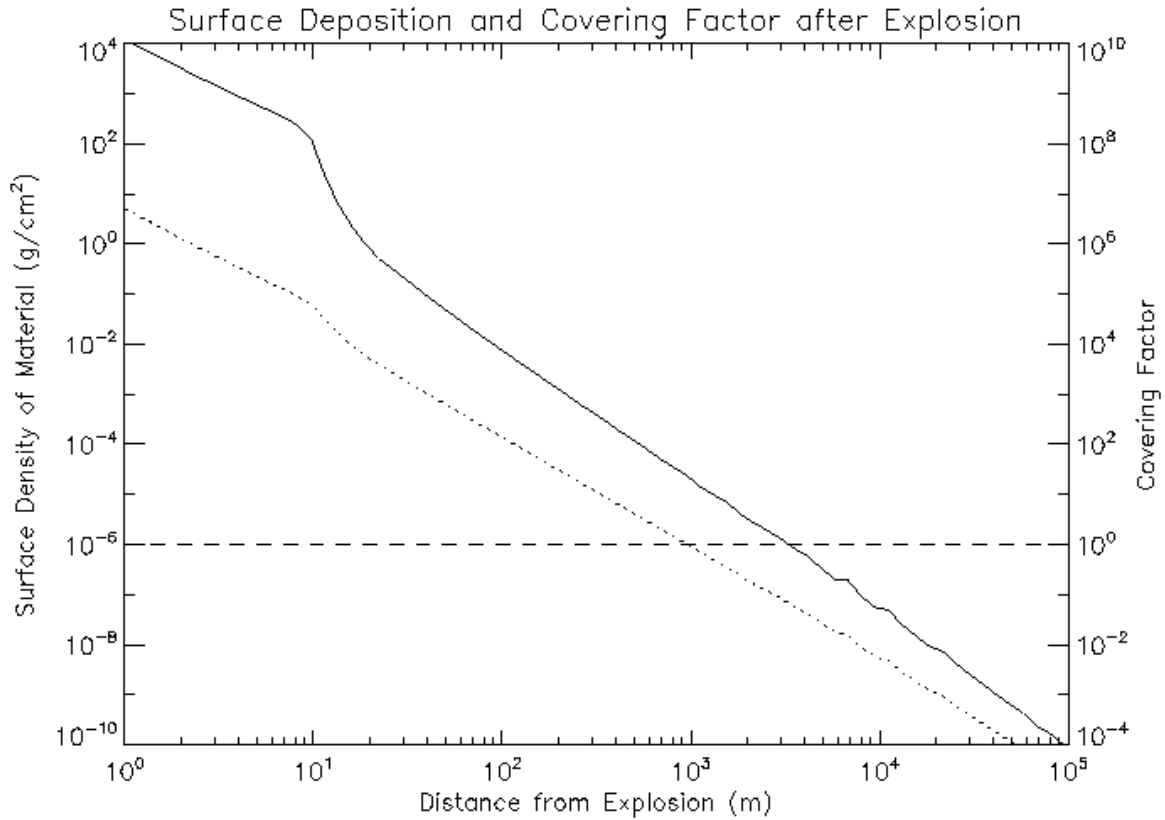}
\caption{
Fig. 3.— The surface density and covering factor of material deposited 
by our model “minimal TLP” versus the distance from the explosion 
center. The surface density (mass per unit area) is the solid curve and 
covering factor (total projected particle area per unit area) is dotted, 
with unit coverage denoted by the dashed line. Away from the crater, 
there is a near power-law of exponent -2.7 for the surface density of 
regolith deposited as a function of radius, whereas for covering factor 
the power-law index is -2.3. Results interior to $r=10$~m are 
complicated by the finite size of the crater and much of the large 
particle material that merely falls near the hypothesized initial 
position. The ratio of these two curves gives the mass-weighted average 
particle size, varying as a power law from 3 $\mu$m at 20 m radius to 0.04
$\mu$m at 100 km.
}
\end{figure}

\begin{figure}
\plotone{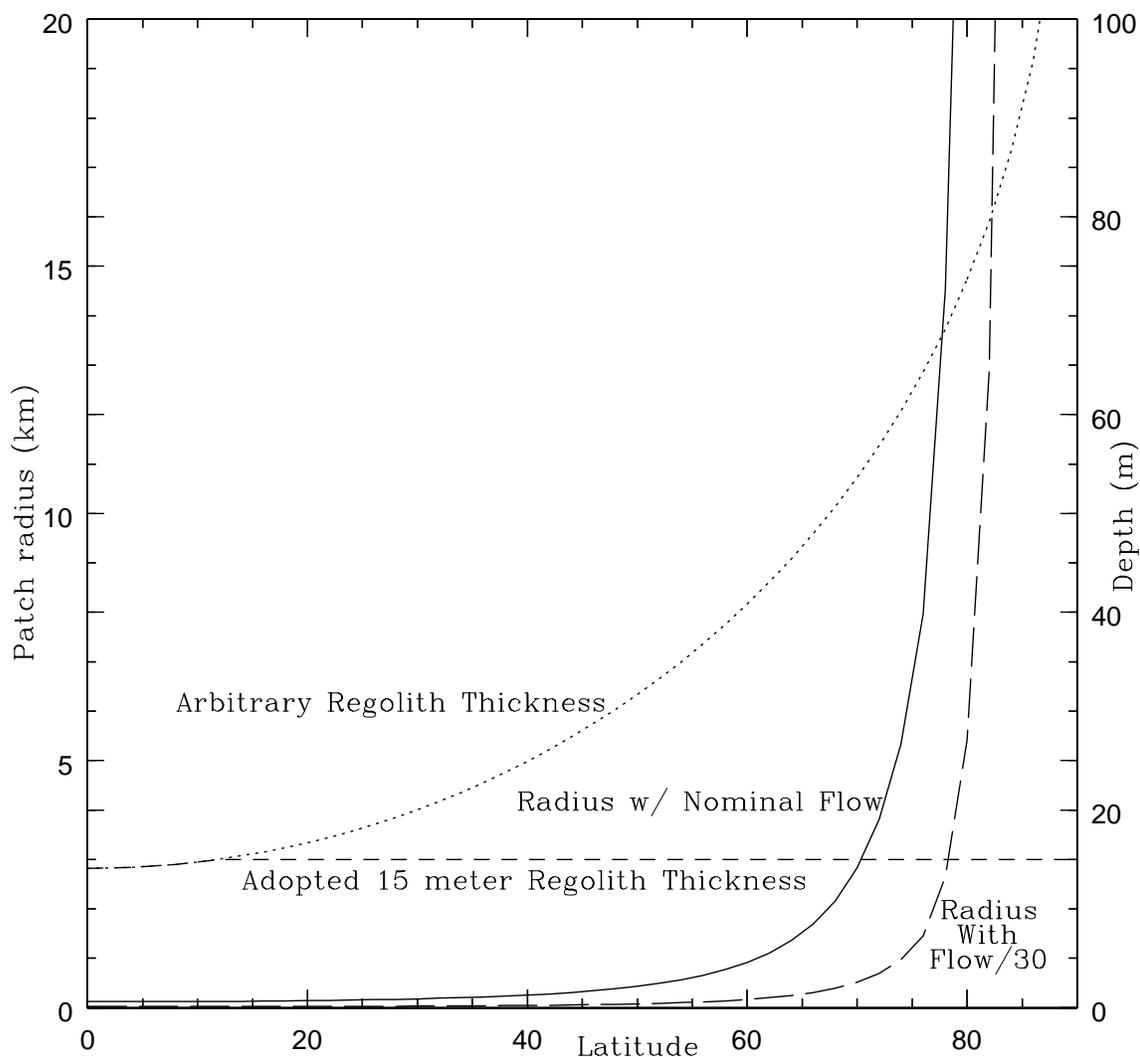}
\caption{
The radius of subsurface ice patches and the maximum depth at which we should
expect growth of ice, as a function of latitude.
The case described in the text corresponds to the solid curve (for ice patch
radius - left axis) and the short dashed curve (for ice depth - right axis).
The arbitrary 15~m limit is adopted assuming that the low diffusivity regolith
overlays a higher diffusivity megaregolith which discourages the growth of ice.
If the regolith is actually deeper, or if an ice cap might actually encourage
growth of ice at greater depth, in principle the ice layer could extend to the
dotted curve (reading the right axis), this would likely encourage the growth
of more ice at a given flow rate.
(If the regolith were surprisingly deep, the ice patch area might grow larger
by a factor roughly the ratio of the dotted curve to the dashed curve.)~
The long-dashed curve is similar to the solid curve, simply showing the size of
the ice patch if the flow rate is reduced by a factor of 30 (to
0.0033~g~s$^{-1}$ of water).
This curve does not account for time required to reach the equilibrium radius.
Smaller than this flow, ice patches might not grow near the equator.
}
\end{figure}

\begin{figure}
\plotfiddle{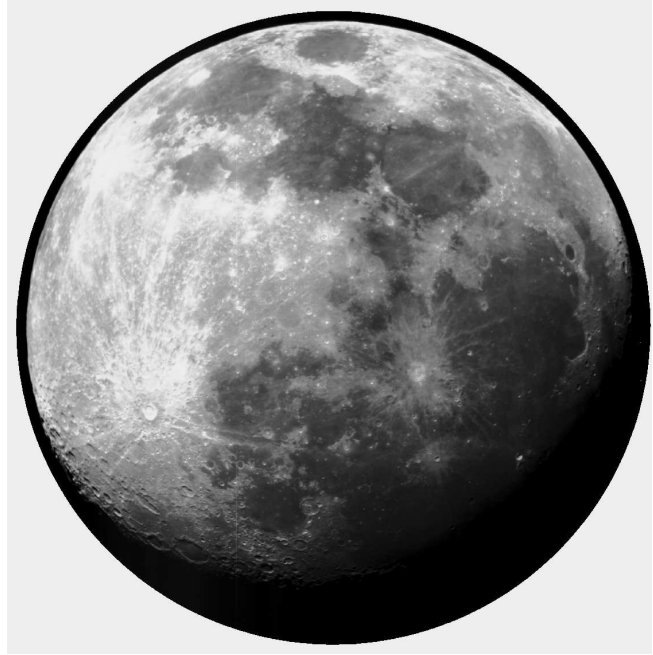}{3.0in}{000}{045}{045}{-127}{-060}
\vskip 0.00in
\caption {
Top)
Flat-field, dark-corrected but otherwise ``raw'' image of a typical lunar Near
Side image obtained by our robotic imaging monitor.
(The image is trimmed to a standard circular region.)~
Bottom)
The difference in signal between the image above and
similar one obtained five minutes later.
The noise in the residual signal is essentially at the photon shot-noise limit.
Because of a slight error in the photometric calibration between the two images
there is a very slight ghost of high-contrast global features, especially 
Imbrium, Humorum and the eastern maria.
Note that even bright smaller features e.g., Tycho left and below center, are
subtracted nearly identically.
Even subtle features not apparent in the image above e.g., an image column
acting slightly non-linearly, just left of center, becomes readily apparent.
There are also errors along the lunar limb due to the rapid gradient in signal
level.
}
\end{figure}

\begin{figure}
\plotfiddle{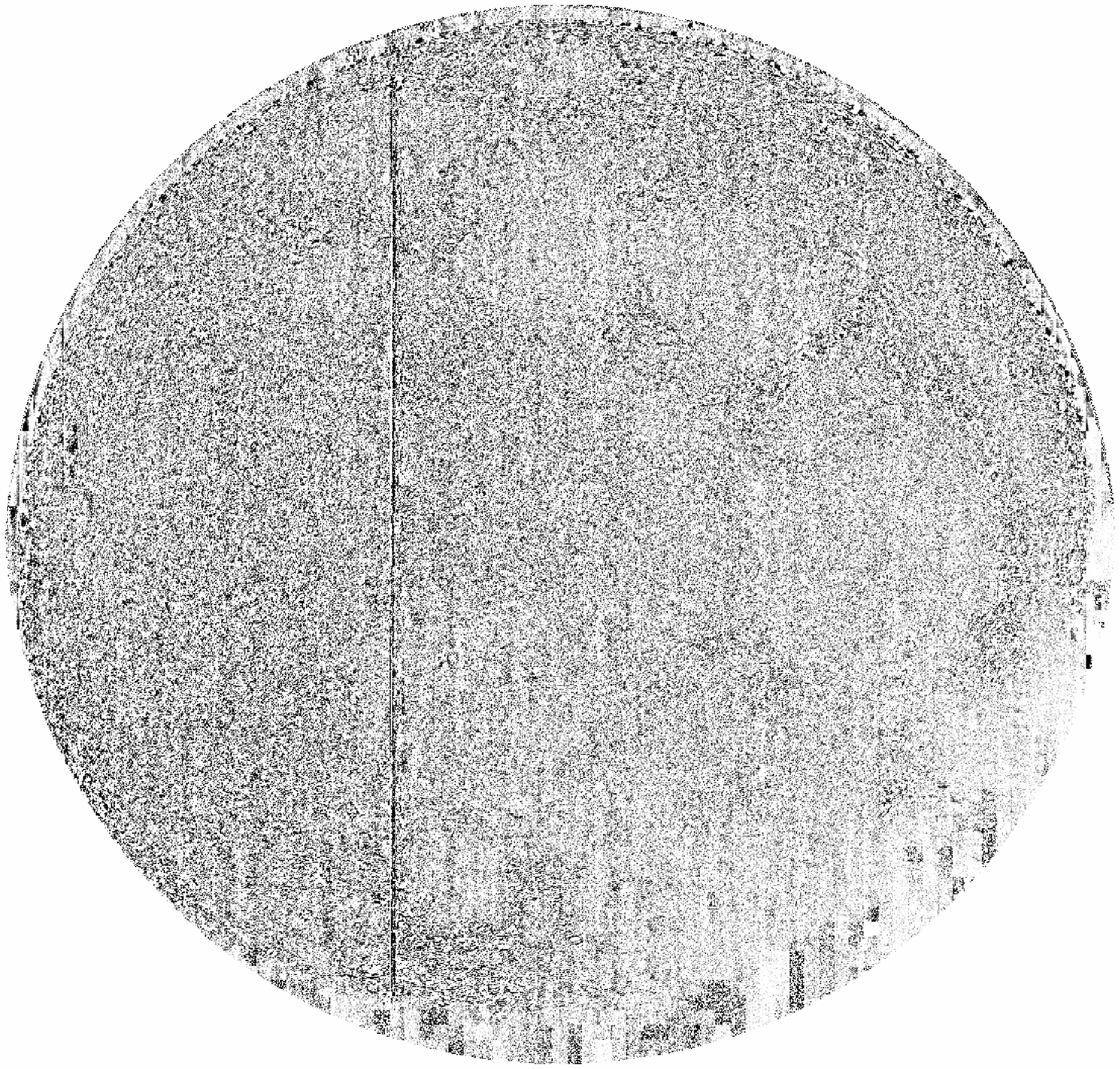}{1.8in}{000}{045}{045}{-127}{-130}
\end{figure}

\begin{figure}
\plotfiddle{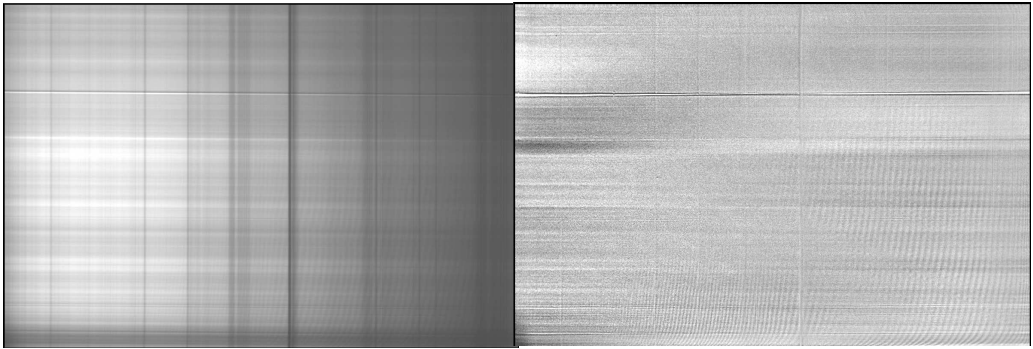}{2.0in}{-90}{030}{030}{-138}{ 683}
\vskip 0.00in
\caption {
{\bf a)} Left:
spectrum of an 8-arcmin slit intersecting Aristarchus (bright streak just above
center) and extending over Oceanus Procellarum, and covering wavelengths
5500-10500\AA, taken by the MDM 2.4-meter telescope;
{\bf b)} Right: the residual spectrum once a model consisting of the outer
product the one-dimensional average spectrum from Figure 7a times the
one-dimensional albedo profile from Figure 7a.
The different spectral reflectance of material around Aristarchus is apparent
(at a level of about 7\% of the initial signal), with r.m.s.\ deviations of
about 0.5\%, dominated by interference fringing in the reddest portion, which
can be reduced.
}
\end{figure}

\begin{figure}
\plotfiddle{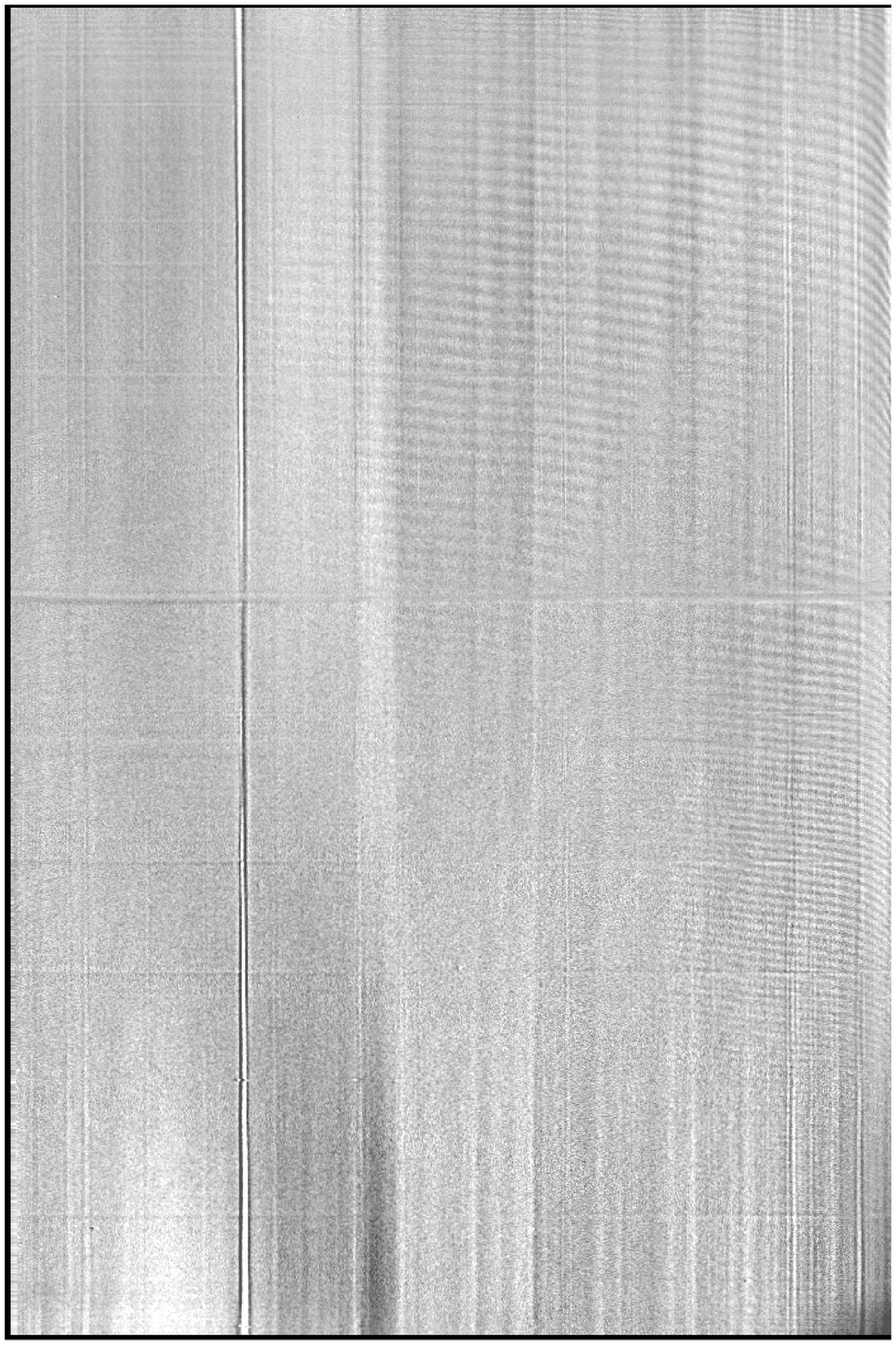}{0.0in}{-90}{030.0}{030.0}{-523}{ 849  }
\end{figure}

\begin{figure}
\plotfiddle{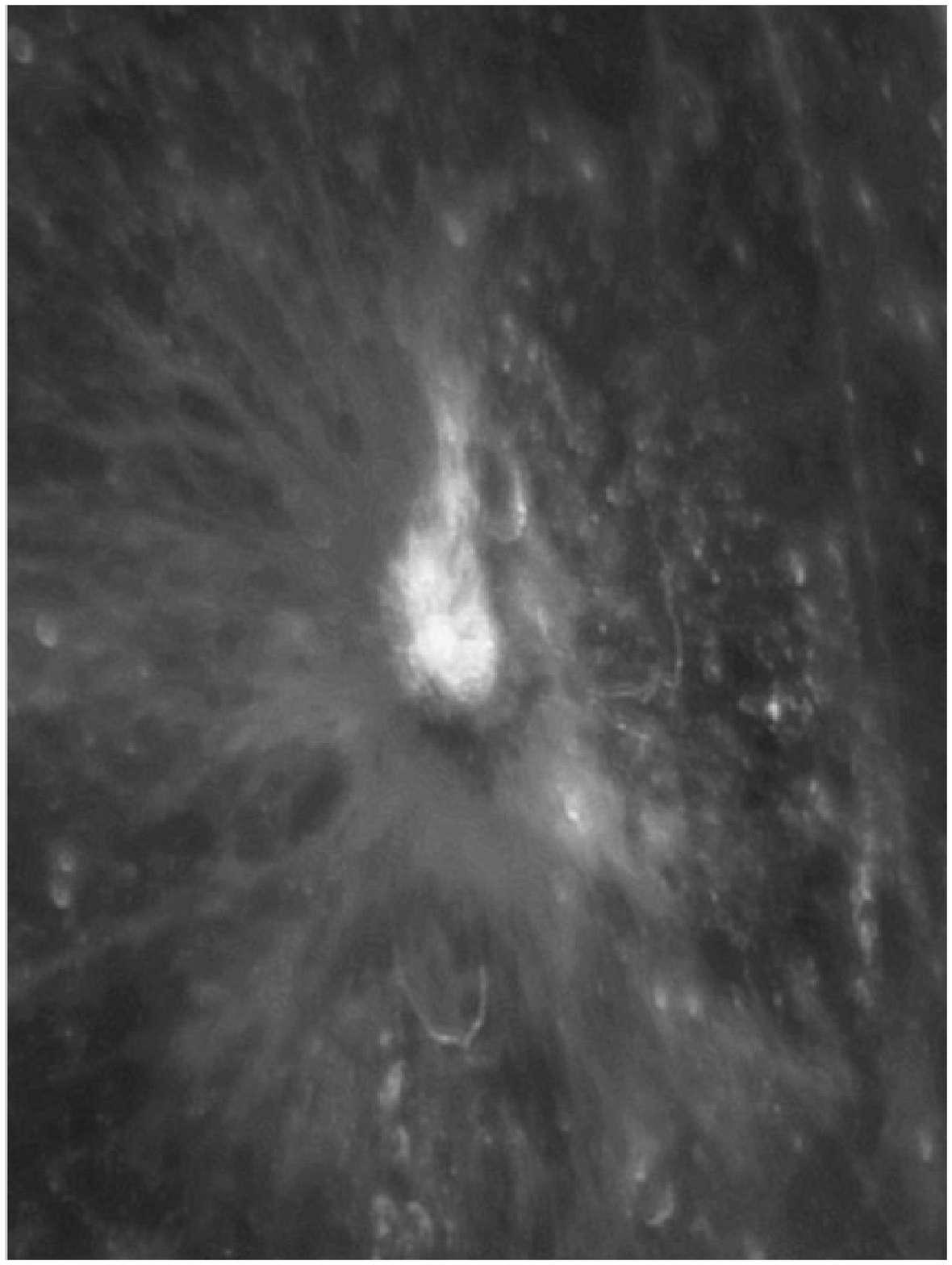}{0.0in}{-90}{050}{050}{-157}{ 200.0}
\end{figure}

\begin{figure}
\plotfiddle{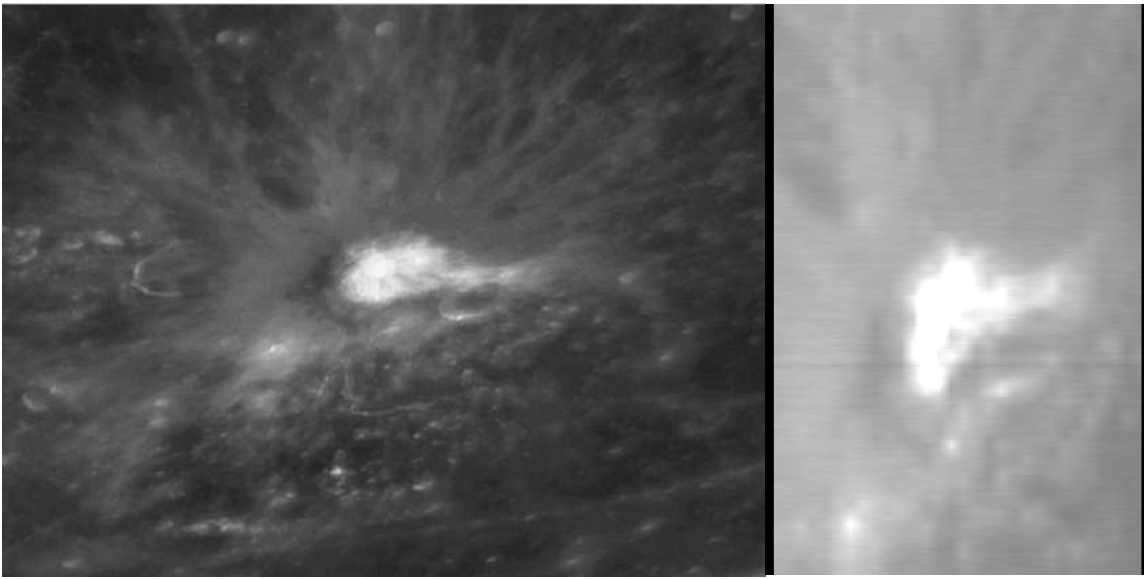}{1.95in}{180}{033}{033}{ 587}{ 889.5}
\vskip 0.00in
\caption {
{\bf a)} Left: a B-band image of the region around Aristarchus;
{\bf b)} Right: an image of Aristarchus in a 3\AA-wide centered near 6000\AA,
constructed by taking a vertical slice through Figure 7a and other exposures
from the same sequence of spectra scanning the surface.
Any such band between 5500\AA\ and 10500\AA\ can be constructed in the same
manner, with resolution of about 1km and 3\AA.
}
\end{figure}

\begin{center}
\begin{deluxetable}{lllll}
\tabletypesize{\scriptsize}
\rotate
\tablecaption{Summary of Basic Experimental/Observational Techniques Detailed Here}
\tablewidth{0pt}
\tablehead{
\colhead{Goal} &
\colhead{Detection Method} &
\colhead{Channel} &
\colhead{Advantages} &
\colhead{Difficulties}
}
\startdata

Map of TLP activity   & Imaging monitor, entire nearside, & optical & comprehensive
schedulability; more & limited resolution  \\
 & $\sim$2 km resolution. & & sensitive than human eye & \\
 & & & & \\
Polarimetric study of & Compare reflectivity in two       & optical & easy to
schedule; further constrains& requires use of two monitors\\
dust                  & monitors with perpendicular       &         & dust behavior 
                    &                     \\
                      & polarizers                        &         &               
                    &                     \\
 & & & & \\
Changes in small,     & Adaptive optic imaging, $\sim$100 m&0.95$\mu$m, etc.&``on
demand'' given good conditions&undemonstrated, depends on \\
active areas          & resolution                        &         &               
                    & seeing; covers $\sim$50 km\\
                      &                                   &         &               
                    & diameter maximum     \\
 & & & & \\
                      &``Lucky Imaging,'' $\sim$200m      &0.95$\mu$m, etc.&on
demand given good conditions& low duty cycle, depends on\\
                      & resolution                        &         &               
                    & seeing               \\
 & & & & \\
                      &{\it Hubble Space Telescope}, $\sim$100 m&0.95$\mu$m, etc.&on
demand given advanced notice& limited availability; low\\
                      & resolution                        &         &               
                    & efficiency           \\
 & & & & \\
                      &{\it Clementine/LRO/Chandrayaan-1} &0.95$\mu$m, etc.&existing
or planned survey   & limited epochs; low flexibility\\
                      & imaging, $\sim$100 m resolution   &         &               
                    &                      \\
 & & & & \\
                      & {\it LRO/Kaguya/Chang\'e-1} imaging, &0.95$\mu$m,
etc.&existing or planned survey   & limited epochs; low
flexibility\\
                      & higher resolution                 &         &               
                    &                      \\
 & & & & \\
TLP spectrum          & Scanning spectrometer map, plus   & NIR,    & may be best
method to find         & requires alert from TLP image\\
                      & spectra taken during TLP event    & optical & composition \&
TLP mechanism       & monitor; limited to long events\\
                      &                                   &         &               
                    &                       \\
 & & & & \\
Regolith hydration    & NIR hydration bands seen before vs.&2.9, 3.4$\mu$m&directly
probe regolith/water  & requires alert from monitor\\
measurement           & after TLP in NIR imaging          &         & chemistry; may
detect water             & and flexible scheduling\\
 & & & & \\
                      & Scanning spectrometer map, then  &2.9, 3.4$\mu$m&directly
probe regolith/water    & requires alert from monitor\\
                      & spectra taken soon after TLP      &         & chemistry; may
detect water             & and flexible scheduling\\
 & & & & \\
Relationship between  & Simultaneous monitoring: $^{222}$Rn $\alpha$&$^{222}$Rn
$\alpha$ \& &refute/confirm TLP/outgassing&optical monitor only covers\\
TLPs \& outgassing    & particles by $Kaguya$ \& optical TLPs& optical &
correlation; find outgassing loci  & nearside; more monitors better\\
 & & & & \\
Subsurface water ice & Penetrating radar from Earth       &$\sim$430 MHz&directly
find subsurface ice with&ice signal is easily confused\\
                      &                                   &         & existing
technique                 & with others  \\
 & & & & \\
                      & Penetrating radar from lunar orbit&$\sim$300 MHz&better
resolution; deeper than&ice signal is easily confused;\\
                      &                                   &         & neutron or
gamma probes           & more expensive\\
 & & & & \\
                      & Surface radar from lunar orbit    &$>$1 GHz & better
resolution; study TLP site  & redundant with high resolution\\
                      &                                   &         &      surface
changes               &            imaging?   \\
 & & & & \\
High resolution TLP   & Imagers at/near L1, L2 points     & optical & map TLPs with
greater resolution \& & expensive, but could piggyback \\
activity map          & covering entire Moon, at 100 m    &         & sensitivity,
entire Moon           & on communications network \\
                      & resolution                        &         &               
                    &                       \\
 & & & & \\
Comprehensive $^{222}$Rn $\alpha$ & Two $^{222}$Rn $\alpha$ detectors in
polar&$^{222}$Rn $\alpha$&map outgassing events at full&expensive; even better
response\\
particle map          &          orbits 90$^\circ$ apart in longitude &         &
sensitivity                        &          with 4 detectors\\
 & & & & \\
Comprehensive  map of & Two mass spectrometers in adjacent& ions \& & map outgassing
events \& find      & expensive; even better with more\\
outgas components     & polar orbits                      & neutrals& composition   
                    &           detectors\\
\enddata
\tablecomments{In situ, surface experiments: we refer the reader to work in
preparation by AEOLUS collaboration.
All methods are Earth-based remote sensing unless specified otherwise.
Abbreviations used: TLP = transient lunar phenomena, NIR = near infrared}
\end{deluxetable}
\end{center}
\end{document}